\documentclass[a4paper,twocolumn,showpacs,notitlepage,floatfix,nofootinbib,superscriptaddress,pre]{revtex4-1}
\usepackage{graphicx}
\usepackage{amsfonts}
\usepackage{amsmath}
\usepackage{hyperref}
\usepackage{float}
\usepackage{amssymb,amsbsy}
\usepackage{epstopdf}
\usepackage{color}
\usepackage{float}
\usepackage{hhline}

\begin{document}

\title{Optimizing direct laser-driven electron acceleration and energy gain at ELI-NP}

\author{Etele Moln\'ar}
\affiliation{Extreme Light Infrastructure – Nuclear Physics ELI-NP, “Horia Hulubei” National Institute for Physics and Nuclear Engineering, 30 Reactorului Street, RO-077125, Bucharest-Magurele, Romania} 

\author{Dan Stutman}
\affiliation{Extreme Light Infrastructure – Nuclear Physics ELI-NP, “Horia Hulubei” National Institute for Physics and Nuclear Engineering, 30 Reactorului Street, RO-077125, Bucharest-Magurele, Romania} 

\author{Catalin Ticos}
\affiliation{Extreme Light Infrastructure – Nuclear Physics ELI-NP, “Horia Hulubei” National Institute for Physics and Nuclear Engineering, 30 Reactorului Street, RO-077125, Bucharest-Magurele, Romania}

\date{\today }

\begin{abstract}
We study and discuss electron acceleration in vacuum interacting with fundamental
Gaussian pulses using specific parameters relevant for the multi-PW femtosecond 
lasers at ELI-NP.
Taking into account the characteristic properties of both linearly and circularly polarized 
Gaussian beams near focus we have calculated the optimal values of beam waist leading to the 
most energetic electrons for given laser power. The optimal beam waist 
at full width at half maximum correspond to few tens of wavelengths, 
$\Delta w_0=\left\{13,23,41\right\}\lambda_0$, for increasing laser power $P_0 = \left\{0.1,1,10\right\}$ PW. 
Using these optimal values we found an average energy gain of a few MeV and highest-energy 
electrons of about $160$ MeV in full-pulse interactions and in the GeV range 
in case of half-pulse interaction.
\end{abstract}

\maketitle

\section{Introduction}

Laser-matter interaction by ultra-intense and ultra-short lasers 
at the Extreme Light Infrastructure (ELI) are intended to provide 
new horizons and breakthroughs in technological advancements and fundamental physics.

Laser-driven charge acceleration has been proposed and investigated both theoretically 
and experimentally during the last few decades 
\cite{Shimoda_1962,Tajima_1979,Sarachik_1970,Scully_1991,Moore_1995,Malka:1997}. 
Plasma based accelerators are an established method to provide electron beams to GeV energies 
in short pulse laser wakefield interactions \cite{Leemans_2006,Leemans_2014}.
Compared to different plasma based acceleration mechanisms where 
the electrons gain energy from the longitudinal fields of charge-separated plasma electrons and ions, i.e., bubble or ion channel 
\cite{Pukhov_1999,Mourou_2006,Esarey_2009,Arefiev_2012, Arefiev_2016, Arefiev_2019}, 
simple vacuum-based acceleration, or direct laser acceleration (DLA), 
has been little studied recently.
However, vacuum-based acceleration is not only much more straightforward and 
simpler involving a well defined and understood 
physical process, but in some cases of interest it might be even 
favorable due to absence of the plasma.

In this paper, we aim to study and review the characteristics of electron acceleration 
and ponderomotive scattering in vacuum by femtosecond laser pulses of highest intensity achievable today.
In particular, we are interested in determining the dependence of the net kinetic energy gain
in the fundamental Gaussian mode, on various laser parameters such as intensity, beam waist, 
pulse duration and polarization.
Our primary purpose is not strictly academic, but we aim to estimate and 
explore the relevant parameters for laser-driven 
electron acceleration for the future experiments at ELI-NP \cite{Tanaka_2020}.

DLA with femtosecond laser pulses and intensities above $10^{19}$ $\textrm{W/cm}^2$, 
were reported to lead to electrons with kinetic energy of a few MeV \cite{He_2003,He_2004}. 
Similarly, here we will study the energy gain of electrons in lasers of intensity, 
above $10^{22}$ $\textrm{W/cm}^2$, operating at ELI-NP.
A possible application of direct electron acceleration is to measure the peak laser 
intensity in the focal region \cite{Kalashnikov_2015}.

Due to the fact that this basic problem in electrodynamics has no known general solution,
the interaction of free electrons with laser fields in vacuum 
is based on a 3-dimensional numerical code that solves the 
relativistic equations of motion. 
In addition to a large number of theoretical papers, e.g., Refs. \cite{Hartemann_1995,Esarey_1995,Quesnel_1998,Hartemann_1998,Stupakov_2000,
Wang_2000,Maltsev_2002,Salamin_2002,
Dodin_2003,He_2005,Salamin_2006,Gupta_2007,Galkin_2008,Harman_2008, Harman_2011,Fortin_2010,
Arefiev_2013,Harjit_2017,Kant_2020},
available in the literature, the novelty in 
our investigation of direct laser acceleration of electrons gives us a good number 
of valuable estimates and insight specifically for the lasers operating at ELI-NP.

The paper is structured as follows. For the sake of completeness 
in Sect. \ref{Lorentz_equations} we recall the relativistic equations of motion 
of charges in electromagnetic fields. 
Next we give a brief summary on the fundamental Gaussian beam \ref{Gaussian_beam} 
in the paraxial approximation. The list of parameters and initial conditions corresponding 
to our study of direct electron acceleration are given in Sect. \ref{initial_conditions}.
Our results and discussions are presented in 
Sect. \ref{Results}. 
Taking into account the characteristic properties of Gaussian beams 
we have calculated the optimal values of beam waist leading to the 
most energetic electrons for given laser power.
Using these values we present and discuss the corresponding electron dynamics and 
energy gains in linearly polarized (LP) and circularly polarized (CP) laser beams.
We found an average energy gain of a few MeV where the highest energy electrons are 
about $160$ MeV, in full-pulse interactions. 
Furthermore in case of half-pulse interaction these energy gains are almost an order of 
magnitude higher, reaching 1 GeV in agreement with the ponderomotive limit.
Finally, the conclusions are summarized in Sect. \ref{Conclusions}.

\section{Direct laser acceleration}
\label{DEA}
\subsection{Charge acceleration in vacuum}
\label{Lorentz_equations}

Here we briefly review the relativistic equations of motion of a charge $q$
in electromagnetic fields \cite{Jackson_Book_1999}. 
The contravariant particle four-momentum $p^{\mu }$ is defined as 
\begin{equation}
p^{\mu }\equiv \left( p^{0},\vec{p}\right) =m_{q}\gamma c\left( 1,\vec{\beta}%
\right),  \label{P_mu}
\end{equation}
where $m_{q}=\sqrt{p^{\mu }p_{\mu }/c^{2}}$ is the rest mass of the charged
particle, $\vec{p}/p^{0}=\vec{v}/c \equiv \vec{\beta}$ is the normalized
three-velocity of the particle and $c=1/\sqrt{\epsilon _{0}\mu _{0}}$ is the speed
of light in vacuum. The Lorentz factor is defined as
\begin{equation}
\gamma \equiv \frac{1}{\sqrt{1-\beta^{2}}}
=\sqrt{1+\left( \frac{|\vec{p}|}{m_{q}c}\right) ^{2}},  \label{gamma}
\end{equation}
where $\beta = |\vec{\beta}|$, while the relativistic three-momentum and the energy 
of the particle are, 
$\vec{p}=\gamma m_{q}c\vec{\beta}$ and $p^{0}\equiv E^{\gamma}/c=\gamma m_{q}c$.

The relativistic equation of motion of charged particles in electromagnetic field is 
determined by the Lorentz force, 
$\vec{F}_L \equiv d\vec{p}/dt = q \left(\vec{E} + \vec{v}\times \vec{B}\right)$. 
This leads to the following set of coupled ordinary differential equations,
\begin{eqnarray}
\frac{d\vec{x}}{dt} &=&c\vec{\beta},  \label{dx_dt} \\
\frac{d\vec{\beta}}{dt} &=&\frac{q}{\gamma m_{q}c}\left[ -\vec{\beta}\left( 
\vec{\beta}\cdot \vec{E}\right) + \vec{E}+c\vec{\beta}\times \vec{B}\right] \, ,
\label{dbeta_dt}
\end{eqnarray}
where $\vec{x}$ denotes the position of the particle, while $\vec{E}$
and $\vec{B}$ are the electric and magnetic fields in vacuum. 
Using the relation from Eq. (\ref{gamma}) together with 
the three independent equations (\ref{dbeta_dt}), provides and additional equation 
for the relativistic energy $E^{\gamma} = \gamma m_{q}c^2$, 
\begin{equation}
\frac{d\gamma }{dt}=\frac{q}{m_{q}c}\left( \vec{\beta}\cdot \vec{E}\right) .
\label{dgamma_dt}
\end{equation}
These equations may be solved in dimensionless form that is found either by
setting all fundamental constants to be one, i.e., $|q|=c=m_{q}=1$, as chosen by us, 
or equivalently by introducing new dimensionless variables, 
i.e., $t_{d}=t\omega _{0}$, $\vec{x}_{d}=\vec{x}\omega _{0}/c$, 
$\vec{p}_{d}=\vec{p}/\left(m_{q}c\right)$, 
where $\omega_{0}$ is the angular frequency.

For given initial conditions, the solution of the relativistic Lorentz equations provide
the velocities and the 3-dimensional trajectories of a moving charge as function of time. 
These solutions depend on the electromagnetic field that will be specified 
in the next section.

\subsection{Gaussian beams}
\label{Gaussian_beam}

Possibly the most well known particular solution to the paraxial wave equation is 
the fundamental Gaussian beam \cite{Siegman_Lasers_1986,Goldsmith_book_1998,Quesnel_1998}. 
The general expression for the transverse electric field of a Gaussian beam is,
\begin{align} \label{E_T}
& E_T \left( r,z\right)  = 
E_{0}\exp \left[-ik_{0}z+i\phi_{0}\right] \\
&\times \frac{w_{0}}{w\left(z\right) }\exp \left[-\frac{r^{2}}
{w^{2}\left( z\right) }-i\frac{zr^{2}}{Z_{R}w^{2}\left(z\right) }
+i\arctan\left(\frac{z}{Z_{R}}\right) \right] , \notag 
\end{align}
where $k_{0}=\omega _{0}/c$ is the wavenumber,  $E_{0}$ is the amplitude of the
electric field, $\phi_{0}$ is the constant initial phase shift and 
$r=\sqrt{x^{2}+y^{2}}$ is the radius in the transverse plane measured from the 
longitudinal axis of propagation, i.e., $z$-axis.
The radius of curvature is 
$R_c (z)\equiv \left(z^{2}+Z_{R}^{2}\right)/z=Z_{R}^{2}w^{2}/\left(z w_{0}^{2}\right)$, 
where the Rayleigh range or confocal distance is defined as 
$Z_{R}= \frac{w_{0}^{2}k_{0}}{2}$. 
Furthermore the Gouy phase at $z$ is given as 
$\phi _{G}\left( z\right)= \arctan\left(\frac{z}{Z_{R}}\right) $, 
while the radius of the Gaussian beam is 
\begin{equation}
w\left( z\right) = w_{0}\sqrt{1+\left( \frac{z}{Z_{R}}\right)^{2}},
\label{waist}
\end{equation}%
where the beam radius at focus (minimum spot size) defines the beam
waist radius $w_{0}\equiv w(z=0)$. 
Therefore for a given laser wavelength and polarization the fundamental Gaussian 
beam has only two parameters, the field intensity and beam waist radius, 
such that from the latter all other parameters describing the beam geometry 
are determined. 
The explicit expressions for the electric field components are,
\begin{eqnarray}\label{ExEy_Gauss}
E_{x}\left( r,z\right) & =& \alpha_{x} E_T, \quad 
E_{y}\left( r,z\right) = -i\alpha _{y} E_T, 
\end{eqnarray}
where $\alpha_{x}=\sqrt{\left(1+\alpha _{P}\right)/2}$ 
and $\alpha_{y}=\sqrt{\left(1-\alpha_{P}\right)/2}$ such that 
$\alpha_{P}=1$ or $-1$ in case of linear
polarization along the $x$-axis or $y$-axis respectively. Furthermore 
$\alpha_{P}=0$ for circular polarization and elliptic polarization
otherwise. Therefore in case of circular polarization the field amplitudes
differ by a factor of $1/\sqrt{2}$ from the amplitudes in case of linear polarization.

Assuming harmonic time dependence of the fields 
$\vec{E}\sim \exp \left( i\omega _{0}t\right)$, the longitudinal 
component of the electric field is calculated from $\nabla \cdot \vec{E}= 0$,
\begin{eqnarray} \label{Ez_Gauss} 
E_{z}\left( r,z\right) &= &-\frac{i}{k_{0}}
\left( \frac{\partial E_{x}}{\partial x}+\frac{\partial E_{y}}{\partial y}\right). 
\end{eqnarray}
Similarly the components of the magnetic field follow from Maxwell's
equations $\vec{B}=\frac{i}{ck_{0}}\nabla \times \vec{E}$ and 
$\nabla \cdot \vec{B}=0$. In the paraxial approximation we have \cite{Kawata_2011}
\begin{eqnarray}
B_{x}\left( r,z\right) &=&-\frac{1}{c}E_{y}\left( r,z\right) , \quad 
B_{y}\left( r,z\right) = +\frac{1}{c}E_{x}\left( r,z\right) , \quad
\label{BxBy_Gauss} \\
B_{z}\left( r,z\right) &=& \frac{i}{ck_{0}}\left( \frac{\partial E_{y}}
{\partial x}-\frac{\partial E_{x}}{\partial y}\right) . 
\label{Bz_Gauss}
\end{eqnarray}%

The finite duration of the pulse is taken into account by assuming the widely used 
Gaussian temporal envelope \cite{Arefiev_2013}, 
where $\tau_{0}$ is the duration of the pulse and $z_{F}$ is 
the initial position of the intensity peak. Therefore introducing, 
\begin{equation}
g\left( t,z\right)= \exp \left[ i\omega _{0}t - 
\left( \frac{t-\left(z-z_{F}\right) /c}{\tau _{0}}\right)^{2}\right] ,  \label{time_envelope}
\end{equation}%
the fundamental Gaussian pulse is completely specified by the components of the 
electric and magnetic fields, Eqs. (\ref{ExEy_Gauss}, \ref{Ez_Gauss}) and 
Eqs. (\ref{BxBy_Gauss}, \ref{Bz_Gauss}), multiplied by Eq. (\ref{time_envelope}), 
and finally taking the real part of the expressions,
\begin{eqnarray} 
\vec{E}\left( t,r,z\right) &=&\text{Re}\left[ \vec{E}\left( r,z\right)
g\left( t,z\right) \right] ,  \label{E_full} \\
\vec{B}\left( t,r,z\right) &=&\text{Re}\left[ \vec{B}\left( r,z\right)
g\left( t,z\right) \right] .  \label{B_full}
\end{eqnarray}

\subsection{Initial conditions}
\label{initial_conditions}

\begin{table*}[hbt!]
\begin{tabular}{|c|c|c|c|c|c|c|}
\hline
$P_0 \,[\textrm{PW}]$ & $I_0\,[\textrm{W}/\textrm{cm}^2]$ & 
$w_0=7.8\lambda_0$
& $I_0\,[\textrm{W}/\textrm{cm}^2]$ & $w_0=13.8\lambda_0$ 
& $I_0\,[\textrm{W}/\textrm{cm}^2]$ & $w_0=24.6\lambda_0$ 
\\  \hhline{=|=|=|=|=|=|=}
$0.1$ & $1.63\times 10^{20}$ & $a_0=8.73$  & $5.21\times 10^{19}$ & $a_0=4.94$ 	&  $1.64\times 10^{19}$& $a_0=2.77$ \\ \hline
$1$   & $1.63\times 10^{21}$ & $a_0=27.63$ & $5.21\times 10^{20}$ & $a_0=15.62$ & 
$1.64\times 10^{20}$& $a_0=8.76$ \\ \hline
$10$  & $1.63\times 10^{22}$ & $a_0=87.37$ & $5.21\times 10^{21}$ & $a_0=49.4$ & 
$1.64\times 10^{21}$& $a_0=27.7$ \\ \hline
\end{tabular}
\caption{The values of peak intensity and normalized field amplitude
corresponding to different waist radii and laser powers.}
\label{Power_table}
\end{table*}

In this paper we study direct laser acceleration of a flat electron cloud 
interacting with different Gaussian pulses. 
Unless stated otherwise, in the following cases of interest, all electrons are 
initially at rest, i.e., $\vec{\beta}_{0,i}=\vec{0}$ hence $\gamma_{0,i} = 1$,
as well as all initial constant phases are set to zero, i.e., $\phi_{0,i}=0$.
The electron charge will be denoted as, 
$q=-e$, while its rest mass as $ m_{e}=0.511 \, \textrm{MeV}/c^2$. 

Furthermore all electrons are initially located at $z_{i}(t_0)\equiv z_{0,i}=0$ 
on the optical axis and distributed uniformly in the transverse plane on a 2-dimensional 
disk with radius that is twice the beam waist radius, 
i.e., $r_{0}=2w_{0}$.
This guarantees that over $99.9\%$ of the laser energy is contained within this disk.
The electron cloud is made of $N=4000$ randomly
distributed non-interacting electrons in the transverse plane for current purposes.
Note that in all cases, we have used the same seed for the pseudo-random number generator, 
hence the initial distribution scales with the radius of the disk.

The laser pulse duration and spot size are specified by values given 
at Full Width at Half Maximum (FWHM).
Therefore a Gaussian of the form, $\exp(-r^2/\sigma^2)$, like in 
Eqs. (\ref{E_T}) and (\ref{time_envelope}), and hence a laser pulse of
$\Delta \tau_0$ duration and beam waist of $\Delta w_{0}$ specified at FWHM translates as, 
\begin{equation}
\tau_{0}=\frac{\Delta \tau _{0}}{2\sqrt{\ln 2}},\ \ \ 
w_{0}=\frac{\Delta w_{0}}{2\sqrt{\ln 2}}.
\end{equation}

Introducing the normalized electric field amplitude $a_{0}=\frac{eE_{0}}{m_{e}c\omega _{0}}$, 
the peak intensity for a linearly polarized Gaussian beam is 
\begin{equation}
I_{0}\equiv E_{0}^{2}\frac{c\varepsilon _{0}}{2}
=a_{0}^{2}\left( \frac{m_{e}c \omega _{0}}{e}\right)^{2}\frac{c\varepsilon _{0}}{2} 
= 3.856 \times 10^{-9} a_{0}^{2} \omega^2_{0}.
\label{I_0}
\end{equation}
The total power carried by the laser beam is 
$P_0\equiv I_{0}\frac{\pi w_{0}^{2}}{2}=6.057\times 10^{-9} a_{0}^{2} w_{0}^{2} \omega^2_{0}$ 
and hence the value of $a_{0}$ for a given power and initial waist radius $w_{0}$ 
is obtained from
\begin{equation}
a_{0}\equiv 1.285\times 10^{4}\ \frac{\sqrt{P_0}}{w_{0} \omega_{0}} 
= 1.61\times 10^{4}\ \frac{\sqrt{I_0}}{\omega_{0}}.  \label{a_0}
\end{equation}%
The value of $a_{0}$ defined above corresponds to a LP laser $a_{0,LP}=a_{0}$, while
in case of a CP laser we have $a_{0,CP}=\frac{1}{\sqrt{2}}a_{0}$.
Thus for a monochromatic laser with wavelength $\lambda_{0}=800$ nm, for any given laser power 
and waist radius $w_{0}$, the normalized field intensities $a_{0}$ are determined 
from Eq. (\ref{a_0}). 

For our cases of interest choosing, $\Delta w_{0}=13\lambda _{0}$, $\Delta w_{0}=23\lambda _{0}$ 
and $\Delta w_{0}=41\lambda _{0}$, the corresponding peak intensity values are listed for 
a linearly polarized Gaussian pulse in Table \ref{Power_table}. 
The outcome of these different cases will be discussed in the next sections.

\section{Results for the fundamental Gaussian beam}
\label{Results}

\begin{figure*}[hbt!]
\vspace{-0.2cm} 
\includegraphics[width=16.4cm, height=4.5cm]
{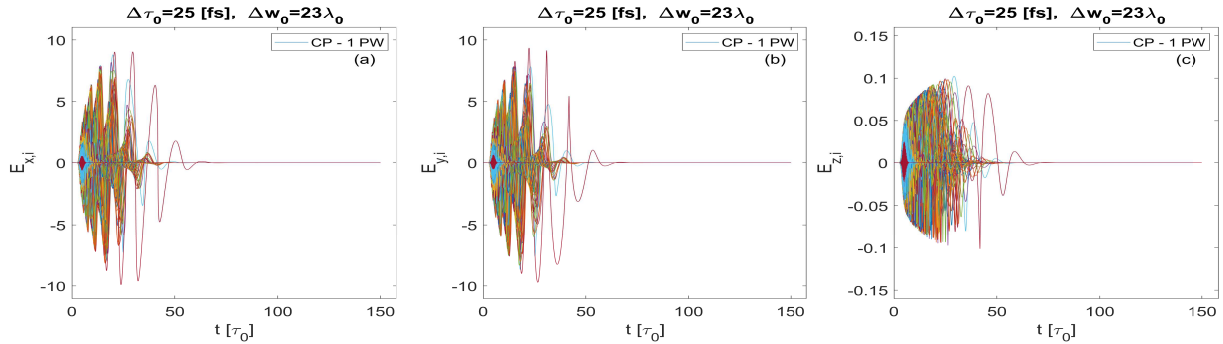} 
\caption{From left to right in $(a)$, $(b)$ and $(c)$ respectively: The Cartesian components of
the electric field, $E_{x,i}$, $E_{y,i}$, $E_{z,i}$, as seen by the charges in units of the normalized 
field intensity $a_0$, as a function of time normalized by the pulse duration. 
The parameters of the Gaussian laser pulse are shown on the figure.}
\label{fig:Efields_Gauss00}
\end{figure*}
\begin{figure*}[hbt!]
\vspace{-0.2cm} 
\includegraphics[width=16.4cm, height=4.5cm]
{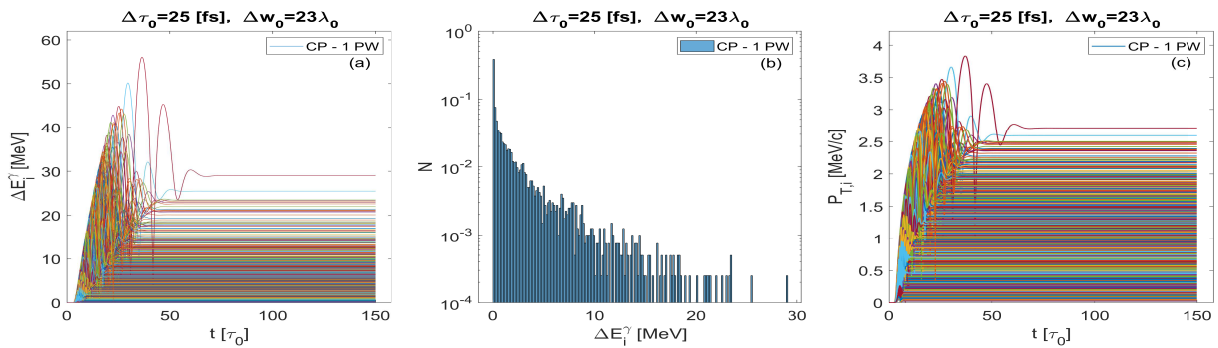}
\caption{From left to right: (a) The net energy
gain of electrons, $\Delta E^{\gamma}_i$, as a function of time, and the corresponding histogram (b), 
followed by the transverse momentum $P_{T,i}$ of electrons (c).
The parameters of the laser pulse are shown on the figure.}
\label{fig:Energy_Gauss00}
\end{figure*}

\begin{figure*}[hbt!]
\vspace{-0.2cm} 
\includegraphics[width=16.4cm, height=4.5cm]
{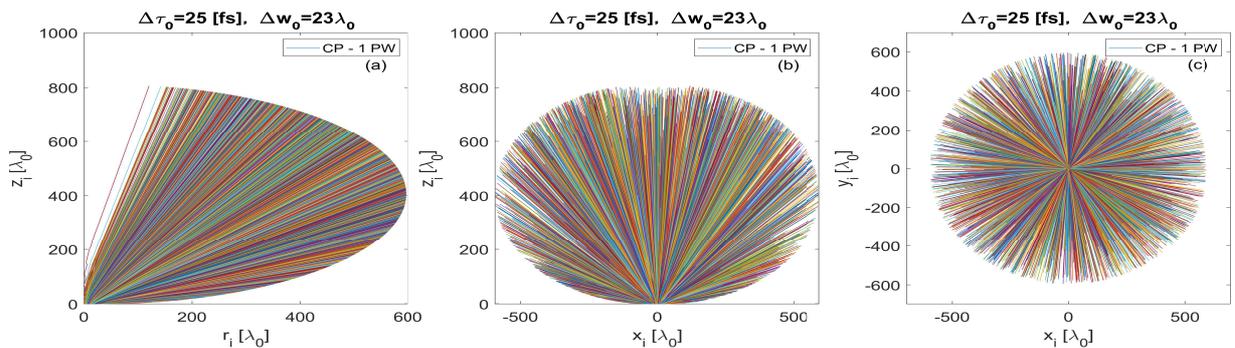}
\caption{From left to right:
(a) and (b) The longitudinal trajectories $z_i$, as function of the radial distance $r_i$, 
and as a function of $x_i$. (c) The electron trajectories in the transverse plane 
$\left(x_i,y_i\right)$. 
The parameters of the laser pulse are shown on the figure.}
\label{fig:XYZ_Gauss00}
\end{figure*}

Here we present and discuss in detail the 3-dimensional numerical 
solutions to particle trajectories and velocities and the 
space-time evolution of the laser pulse.
These solutions are obtained by solving the equations of motion 
Eqs. (\ref{dx_dt}, \ref{dbeta_dt}) together with 
the electromagnetic field of the propagating laser pulse Eqs. (\ref{E_full}, \ref{B_full}) 
for a large number of independent and randomly distributed electrons.
The numerical solutions are required to have accuracy and numerical precision 
up to 12-digits, by using an adaptive time-step Runge-Kutta method available 
in programming languages like Mathematica or Matlab.

The physics of laser-driven electron acceleration in vacuum follows from the 
direct interaction of the laser pulse with electrons as given by the Lorentz force,
Eqs. (\ref{dx_dt},\ref{dbeta_dt}).
For $a_0\geq 1$, the energy and momentum oscillations induced by the laser become relativistic,
and the Lorentz force converts a large part of the laser's energy into longitudinal momentum. 

Any electric charge interacting with the laser pulse is accelerated 
by the electric and magnetic fields therein.
Therefore a point-like charge gains momentum and will get displaced from 
its current location to a new position, where the electric
and magnetic fields of the laser pulse are in general different from before 
(one time-step earlier). 
In this way, the propagating electric charge dynamically maps the laser pulse's 
electric and magnetic fields along its trajectory. 
To explain it in another way, since the electromagnetic field is function of both space 
and time, therefore at any given time the particle trajectories, i.e., the solutions to 
Eqs. (\ref{dx_dt},\ref{dbeta_dt}), serve as input, space-time coordinates, 
for the electromagnetic field of the pulse, i.e.,
Eqs. (\ref{E_full}, \ref{B_full}). 

Furthermore, the electric field, Eqs. (\ref{ExEy_Gauss}), 
and thus the intensity of the Gaussian pulse is highest at the center 
$x_{0}= y_{0}=z_{0}\equiv 0$, i.e., peak intensity, but decreases 
exponentially with increasing transverse radial distance $r^2$. 
Since the charges are distributed randomly in the transverse
plane the available electric and magnetic fields
are also different in different positions. 
Therefore, in general the net energy gain of electrons interacting with 
a laser pulse is a function of the intensity and a function of the initial position 
of charges in the focus of the laser and the beam waist.
In the next sections, we will discuss these cases in more detail.

\subsection{Full pulse interaction $z_F = -5 \tau_0 c$}
\label{full_pulse}

In this case we assume that the initial position of the peak of the Gaussian laser pulse 
is located on the optical axis behind the charges at $z_{F}=-5\tau _{0}c$. 
This guarantees that the electric field in the front part of the pulse initially acting 
on the electrons is vanishingly small, and therefore the initial constant phases are not 
affecting the trajectories.

A circularly polarized  $P_0 = 1$ PW Gaussian beam of  
$\Delta \tau_0 = 25$ fs pulse duration and focal spot diameter 
of $\Delta w_0 = 23\lambda_0$ at FWHM, was shot on a flat disk of electrons, 
according to the initial conditions summarized in Sect. \ref{initial_conditions}.
Since the laser pulse is initially behind the particles, 
the electrons are only gradually overtaken and accelerated by the propagating pulse. 
In Fig. \ref{fig:Efields_Gauss00} we have plotted the Cartesian
components of the electric field seen by the electrons with lower $i$-index 
as a function of time.
The longitudinal component of the electric field $E_{z,i}$ was
calculated from Eq. (\ref{Ez_Gauss}) leading to almost 2-orders of
magnitude smaller values than the corresponding transverse components of the
electric field \cite{Cicchitelli_1990}. 
In a linearly polarized pulse these longitudinal field components will decelerate 
the electron and reduce the net-energy gain \cite{Quesnel_1998,Maltsev_2002}.
However in a circularly polarized Gaussian pulse the effect is opposite and 
the electrons are accelerated by the longitudinal fields.

The visible asymmetry in the time-evolution
of the electric field as traced by the particles, is partially caused
by the longitudinal components of the $\vec{v} \times \vec{B}$ force 
and of the electric field,
as well as by the temporal envelope of the finite pulse.
The electrons are accelerated to larger and larger velocities 
in the front part of the pulse and thus the electron trajectories 
will become elongated in the direction of the laser propagation. 
Therefore the deceleration in the back part of the pulse becomes less efficient, 
hence the asymmetry.

Also note that the normalized electric field amplitudes corresponding to the peak pulse 
position have intensities that are larger (see the $a_0$ values in Table \ref{Power_table}) 
than accessible to even the highest energy electrons during the interaction with the pulse.
This means that the electrons were scattered out of the pulse earlier 
and without reaching the available peak intensity of the pulse. 

In Fig. \ref{fig:Energy_Gauss00}a the time-evolution of the net energy gain of electrons
\begin{equation}
\Delta E^{\gamma}_{i}\left( t\right) \equiv
E^{\gamma}_{i}\left( t\right) - E^{\gamma}_{i}\left( t_{0}\right) 
= \left(\gamma_i - 1\right)m_e c^2 \, ,
\end{equation}
is shown, where the initial energies are $E^{\gamma}_i (t_0) =  m_e c^2$.  
The corresponding histogram and the evolution of the electrons transverse momentum 
$P_{T,i}= \sqrt{p^2_{x,i} + p^2_{y,i}}$  are also shown in 
Figs. \ref{fig:Energy_Gauss00}b and \ref{fig:Energy_Gauss00}c.
Comparing the gain in energy to the gain in transverse momentum we see that the 
kinetic energy of particles is predominantly contained in the longitudinal momentum, 
i.e., $p_z = \gamma m_e v_z$.

Furthermore, we also observe from the energy histogram that about $40\%$ of electrons gain very little 
or no net energy from the laser pulse. 
This is due to the fact that the initial transverse area where the electrons 
are located is relatively large, i.e., $r_{0}=2w_{0}$, while the intensity 
of the electromagnetic field falls off exponentially with the transverse radius squared. 
Therefore for $r_{0} \leq 1.1 w_{0}$ more than $90\%$ of the laser power is captured 
while the electrons beyond this transverse area are interacting with 
relatively weak fields which results in low energy gains.

The charges initially located around the laser axis,
are pushed out from the center by the radial ponderomotive force. 
The full particle trajectories are shown in Figs. \ref{fig:XYZ_Gauss00}a-\ref{fig:XYZ_Gauss00}c,
a long time after the laser pulse passed.
In Fig. \ref{fig:XYZ_Gauss00}a, we have plotted the longitudinal 
trajectories as a function of the radial distance from the origin.
Similarly, Fig. \ref{fig:XYZ_Gauss00}b shows the longitudinal trajectories 
as function of one of the transverse axes, $x$-axis. 
The plot as a function of the other transverse axis is very similar, therefore not shown. 
Observe that the trajectories in the longitudinal direction are on 
average twice longer than in the transverse directions, meaning that the average 
velocities in the longitudinal direction are also larger than in the transverse directions, 
as already concluded before.

Furthermore, it is also interesting to note that the particles around the center 
have all scattered out with a polar angle of 
$\theta_i = \arccos \left( z_i/r_i\right)\geq 0$.
For a given waist radius this angle decreases with increasing laser power, 
meaning that the particles around the center will gain larger 
longitudinal velocities and travel farther in the longitudinal direction.

\begin{figure}[hbt!]
\vspace{-0.2cm} 
\includegraphics[width=6.cm, height=4.5cm]
{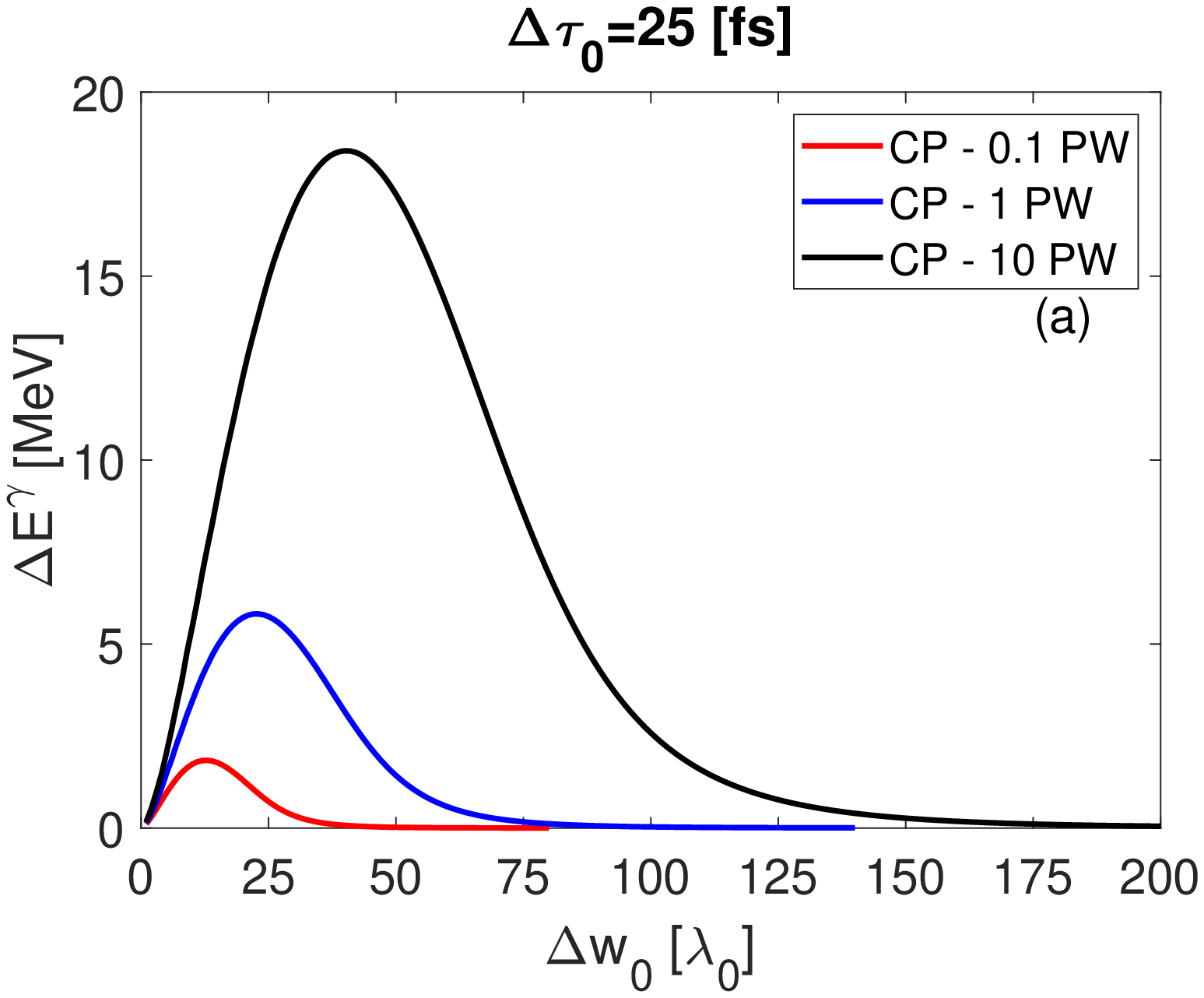}\\
\includegraphics[width=6.cm, height=4.5cm]
{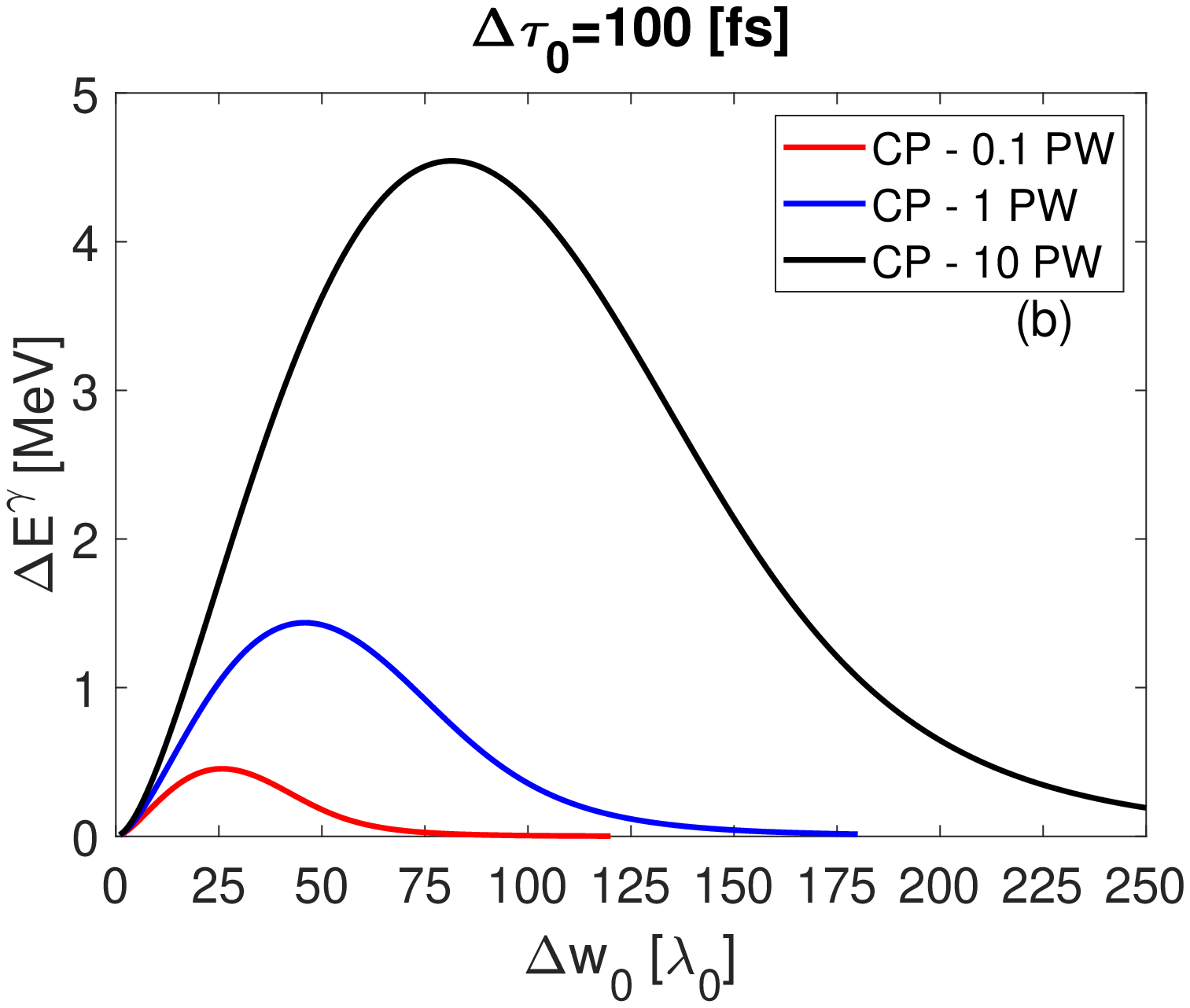}\\
\caption{The average energy gain of an electron initially located at $z_0=0$ 
and $x_{0}=y_{0}\equiv \left\{0.1,0.15,\cdots,1.1 \right\} w_0$ as a function of 
the beam waist. The red, blue and black lines correspond to the weighted average energy for 
laser powers of $P_0=\left\{0.1, 1, 10\right\}$ PW respectively.
The laser pulse duration is (a) $\Delta \tau_0 = 25$ fs and (b) $\Delta \tau_0 = 100$ fs 
as shown on the top of the figure. 
These results are practically independent of the polarization, hence only the CP results are shown.}
\label{fig:w0_scan}
\end{figure}
\begin{figure*}[hbt!]
\vspace{-0.2cm} 
\includegraphics[width=16.4cm, height=4.5cm]
{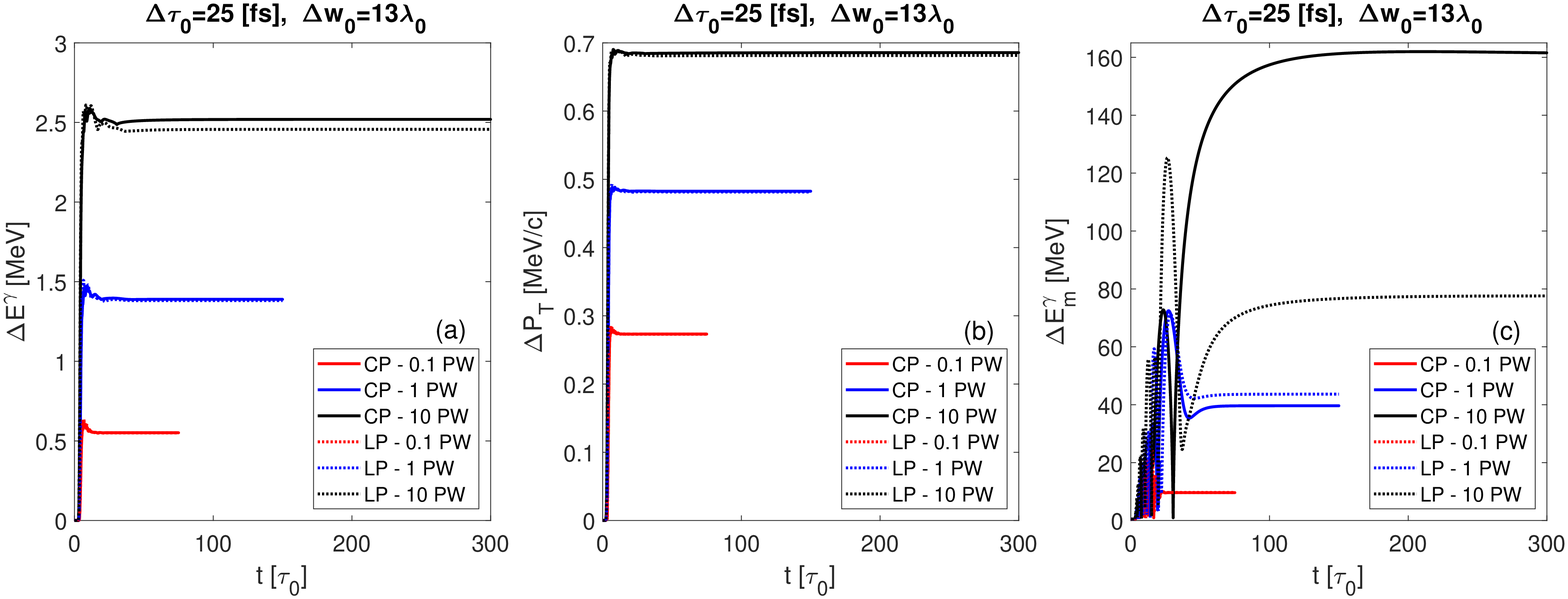} 
\includegraphics[width=16.4cm, height=4.5cm]
{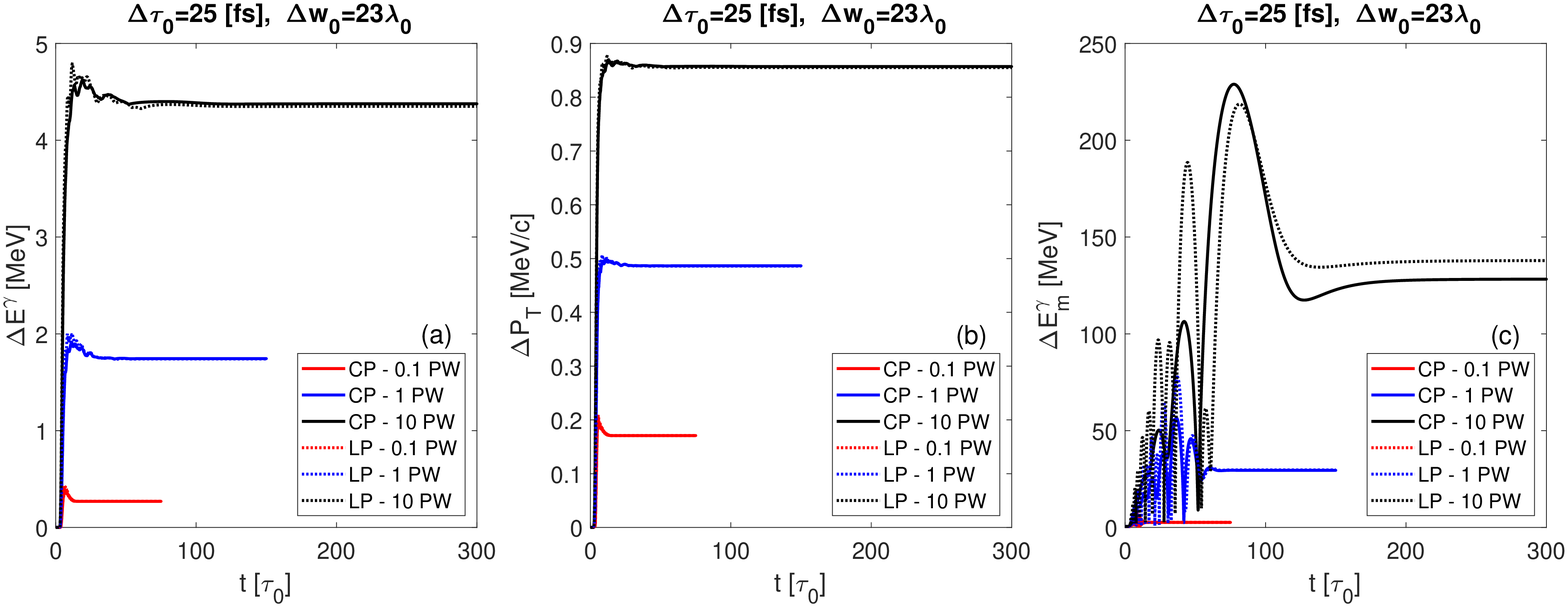} 
\includegraphics[width=16.4cm, height=4.5cm]
{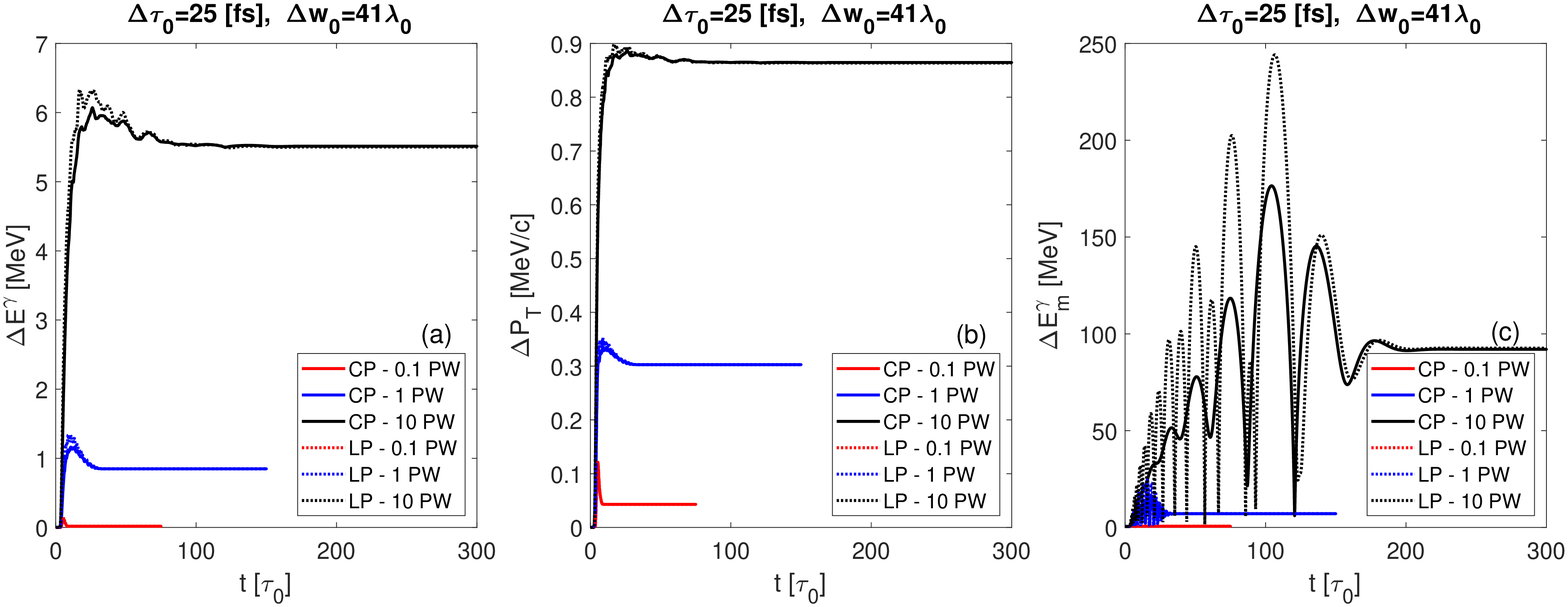}
\caption{Upper row: All figures correspond to $\Delta w_0=13\lambda_0$ beam waist. 
From left to right: $(a)$, $(b)$ and $(c)$ respectively. 
The time evolution of the mean net energy gain of electrons for laser powers 
of $P_0 = \left\{0.1, 1, 10\right\}$ PW, 
with red, blue and black correspondingly. The full and dotted lines 
are for circularly and linearly polarized pulses respectively.
Similarly, (b) shows the average transverse momentum gained from the pulse as 
a function of time.
The rightmost figure (c), shows the time evolution of the highest energy electron, $\Delta E^{\gamma}_m$, 
corresponding to given laser power. 
Middle row: The same as the before, but for an initial beam waist of $\Delta w_0=23\lambda_0$.
Lower row: The same as the before, but for an initial beam waist of $\Delta w_0=41\lambda_0$.
}
\label{fig:Energy_CPLP}
\end{figure*}
\begin{figure}[hbt!]
\vspace{-0.2cm} 
\includegraphics[width=8.2cm, height=4.5cm]
{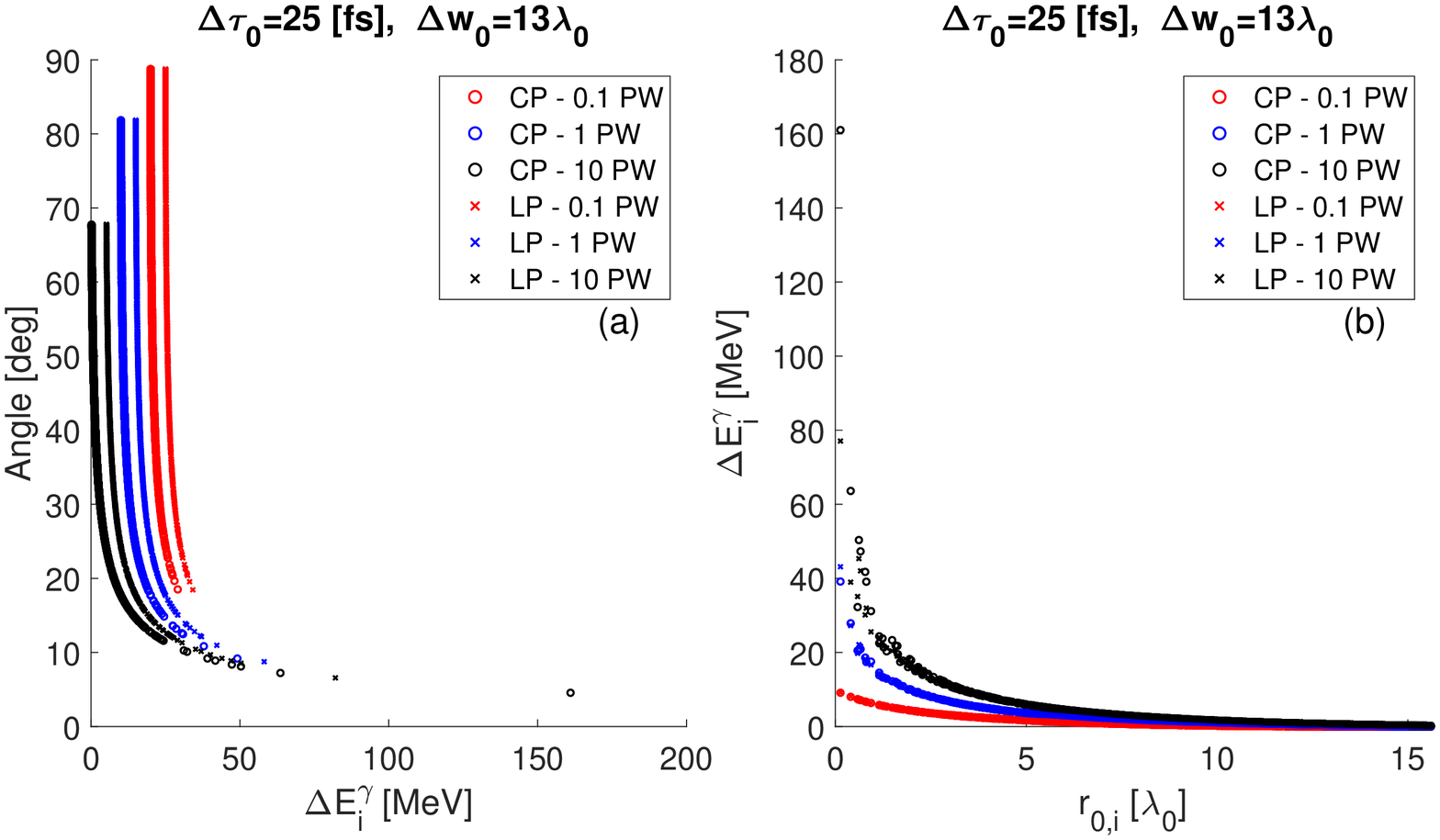} \\
\includegraphics[width=8.2cm, height=4.5cm]
{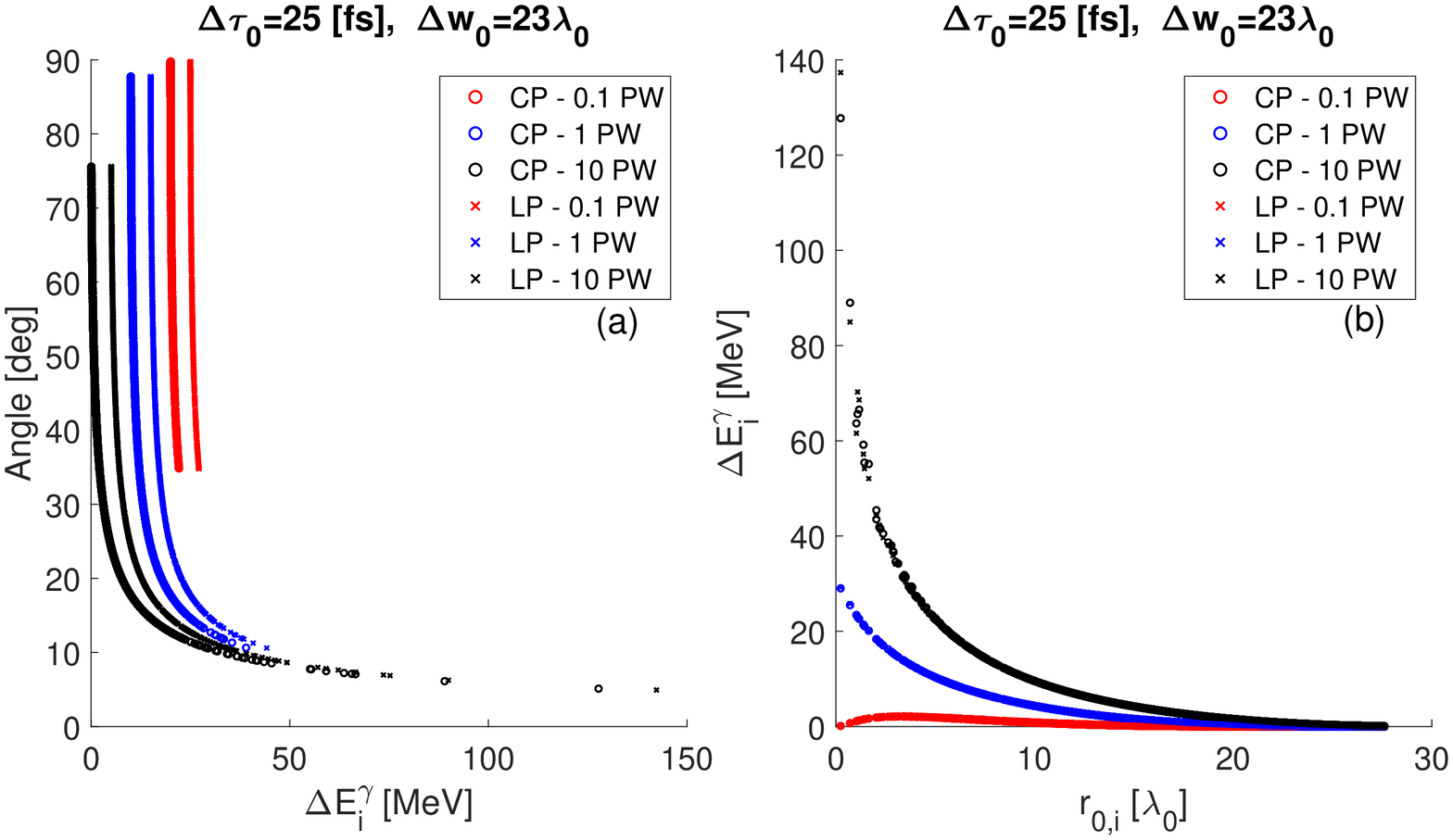} \\
\includegraphics[width=8.2cm, height=4.5cm]
{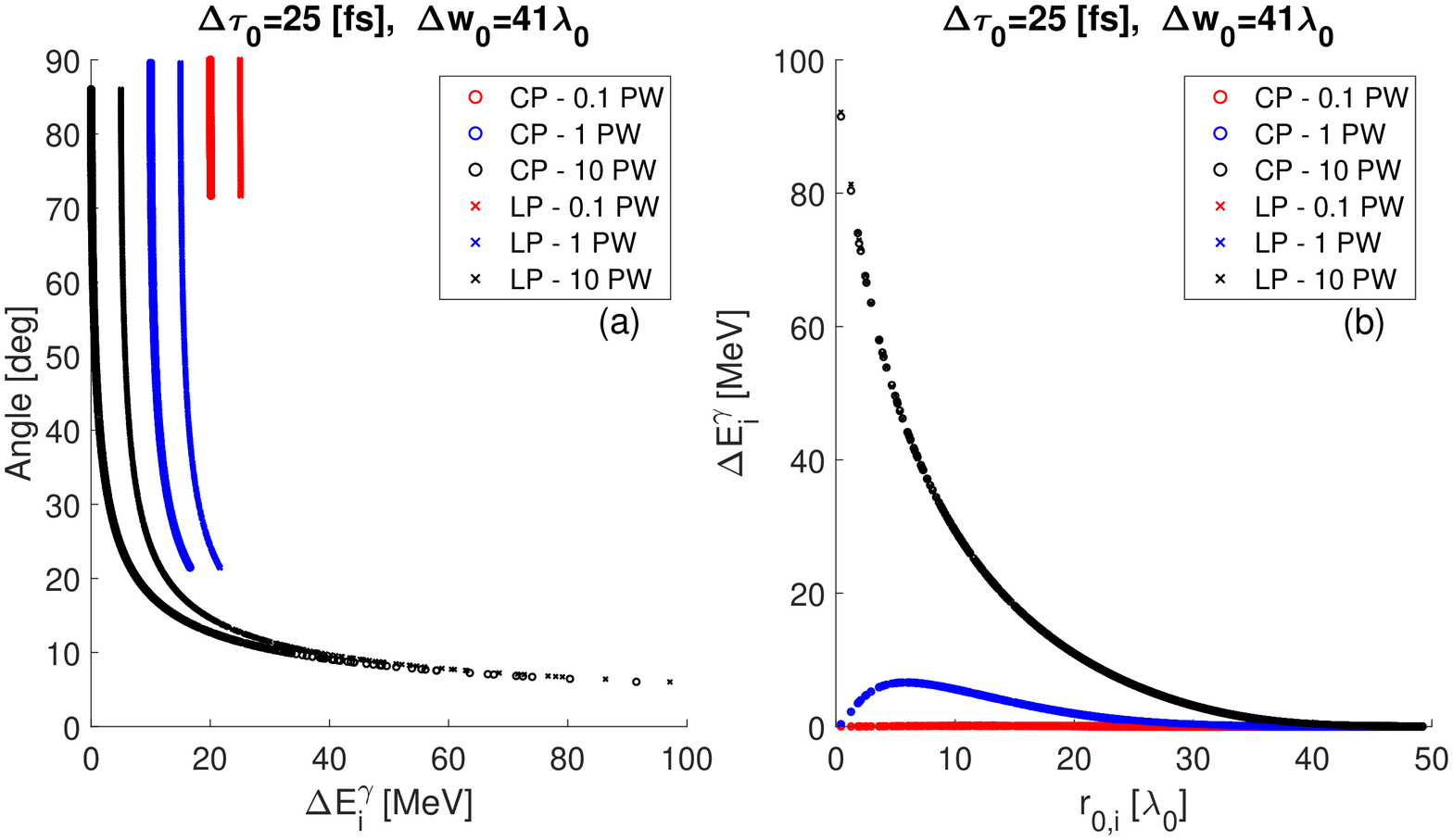}
\caption{From left to right (a) and (b): The figure on the left shows the scattering angle calculated from Eq. (\ref{angle_energy}) as a function of energy gain. The data points corresponding to the CP and LP pulse are plotted with "o" and "x". For better visibility, otherwise the data points they lay on the same curve, 
the data corresponding to a CP pulse with $0.1$ PW and $1$ PW, red and blue, 
are shifted by $10$ and $20$ MeV. 
Similarly for the LP pulses the points are shifted to the right by $5,15,25$ MeV.
(b) The figure on the right shows the energy gained form the pulse as a function of 
the initial radial distribution of electrons. The CP and LP results are basically indistinguishable. 
Middle row: The same as before, but for an initial beam waist of $\Delta w_0 = 23 \lambda$.
Lower row: The same as before, but for an initial beam waist of $\Delta w_0 = 41 \lambda$.
}
\label{fig:Angular_CPLP}
\end{figure}

\subsection{Optimizing the waist of Gaussian pulses}
\label{full_pulse_optimized}

The energy gain of charges interacting with the fundamental Gaussian laser pulse is not 
a linear function of the initial location of electrons or the beam waist. 
Here we have estimated the value of the beam waist that correspondingly leads 
to maximum energy gains. 
First, for a given laser power and for all discrete values of 
$\Delta w_0 = \left\{1,2,\cdots,250\right\}\lambda_0$ we have varied the initial 
position of particles such that they are located at given transverse coordinates 
$x_{0}=y_{0}\equiv \left\{0.1,0.15,\cdots,1.1 \right\} w_0$,
and then calculated the respective energy gains.
Then for every given laser power we have calculated the radially weighted average of these results, i.e., $\Delta E^\gamma_r = \sum_i r_i \Delta E^\gamma_i/\sum_i r_i$.
These weighted averages are shown in Fig. \ref{fig:w0_scan}a and Fig. \ref{fig:w0_scan}b 
for different pulse durations and laser powers.
Henceforth for given laser power, we can read the values of beam waist
corresponding to the peak in energy gain. 
For a laser with $\Delta \tau_0=25$ fs, these values are also listed in 
Table \ref{Power_table}, and hence represent the optimal values for direct electron 
acceleration for given laser power at ELI-NP.

Increasing the laser power will increase the net energy gain for a given spot size, 
see Figs. \ref{fig:w0_scan} which allows the comparison of the energy gain at any given waist.
At the same time, for any given laser power, increasing the beam waist also leads to an increase 
in the energy gain. This is because for a wider beam waist, the scattering of 
the electrons decreases and hence they may remain confined in the pulse for longer.
As a consequence they are able to extract more energy and momentum from the laser.
With increasing intensity the amplitude of the electron oscillations 
along the polarization direction increases until it becomes larger than the size of the 
beam, and the electrons scatter out in the transverse direction.

On the other hand, increasing the beam waist leads to an increase in energy gain at first,
but only until the net energy gain reaches a peak or saturation point which corresponds 
to the optimal beam waist for given laser power.
Increasing the waist radius any further than the corresponding peak
value will in fact, lead to a decrease in net energy exchange with the laser. 
This is because, widening the beam waist beyond optimal values will reduce 
the longitudinal components of the electromagnetic forces.
Thus in conclusion for larger laser power a wider but optimal initial waist ensures 
that the electrons remain confined inside the pulse for longer and hence accelerated 
to higher energies.

How much energy and momentum the electrons gain also depends on the duration 
of the laser pulse.
For any given laser power increasing the pulse duration decreases the 
energy gain, as it should be since for a given power, $P_0 \sim \Delta E^{\gamma}/\Delta t$. 
Therefore for optimal energy gain, a longer pulse requires longer time spent inside 
the pulse, which translates to a wider beam waist. 
Reading and comparing the data from Figs. \ref{fig:w0_scan}, for a given laser power 
increasing the pulse duration by a factor of four, requires approximately two times larger 
beam waist radius.

Now, using the optimal values for the beam waist at given 
laser power, we have studied the electron dynamics for laser pulses of 
$\Delta \tau_0 = 25$ fs duration corresponding to the nine different cases listed in Table \ref{Power_table}.
These are shown in Figs. \ref{fig:Energy_CPLP} for circularly and 
linearly polarized pulses.

The rightmost columns of Figs. \ref{fig:Energy_CPLP}
show the time evolution of the average energy gained 
corresponding to different waist sizes.
The energy averages $\Delta E^{\gamma}= \frac{1}{N} \sum_{i=1}^{N} \Delta E^{\gamma}_i$ 
for $N=4000$ electrons from circularly polarized with full lines and  
from linearly polarized Gaussian pulses with dotted lines are shown in 
Figs. \ref{fig:Energy_CPLP}a.
Similarly, Figs. \ref{fig:Energy_CPLP}b show the average transverse momentum, 
$\Delta P_T=\frac{1}{N} \sum_{i=1}^{N} \sqrt{p^2_{x,i} + p^2_{y,i}}$, 
while the leftmost columns in  Figs. \ref{fig:Energy_CPLP}c show the electrons with 
the highest energy gain, $\Delta E^{\gamma}_m$, at given waist size and laser power.

The right columns of Figs. \ref{fig:Angular_CPLP} show the net energy gain 
as a function of the initial transverse radial distance, 
$r_{0,i} \equiv r_i(t_0) = \sqrt{x^2_{0,i} + y^2_{0,i}}$, from 
the origin at $z_{i,0}\equiv z_{i}(t_0)=0$ for varying laser power. 
Furthermore, from the conservation of energy and momentum it follows that, 
the electrons that were initially at rest are ejected out from the pulse 
with an angle of \cite{Moore_1995,Hartemann_1995,Maltsev_2002}, 
\begin{equation}
\theta^{\gamma}_{i} = \arctan \left( \sqrt{ \frac{2 m_e c^2}{\Delta E^{\gamma}_i}}\right). 
\label{angle_energy}
\end{equation}
This so-called ponderomotive scattering angle as a function of energy gain is 
represented by black, blue, and red symbols for given laser power in Figs. \ref{fig:Angular_CPLP}b. 
For example comparing the red data points for the different waist sizes, we see that for 
a given laser power the scattering angles reach the smallest values for the optimal 
beam waist.
Similarly for the highest laser power the scattering angle asymptotically decreases 
as a function of energy.
These curves are insensitive to the polarization plotted with 
"o" and "x" for CP and LP beams, hence for better visibility the blue and red 
data points are shifted by $10,20$ MeV in the CP case compared to the black "o" symbols touching 
the vertical axis. 
Similarly in the LP case the black, blue, and red, "x" symbols are shifted by 
$5,15,25$ MeV.

As previously discussed, the intensity of Gaussian beam falls off exponentially as 
a function of the radius squared, hence the charges found further away from the 
center will gradually gain less and less energy.
This is so because the laser intensity and hence the radial ponderomotive decreases with 
the increase of beam waist and the charges remain confined near the center. 
However, for larger radial distance from the center, the energy gain increases, 
up to a maximum, while further away the electrons will gain less 
energy attributed to the shorter time/length span in the pulse, 
see the blue and red data points in the lower row of Fig. \ref{fig:Angular_CPLP}b.
Therefore the lower energy electrons are more likely to scatter out 
with larger $\theta^{\gamma}$ angles as shown in Figs. \ref{fig:Angular_CPLP}a.

The overall conclusions regarding the energy gain in full
Gaussian pulse interactions are straightforward. 
The average energy gain of electrons increases with increasing laser power, 
and for a given laser power it is largest for an optimal beam waist $\Delta w_0$. 
For example comparing the black lines in the first column of Figs. \ref{fig:Energy_CPLP}, 
corresponding to $P_0=10$ PW power, we see that for 
the highest laser power, the largest mean energy is gained for the largest beam waist. 
Similarly comparing the red lines corresponding to the lowest power laser, i.e., 
$P_0=0.1$ PW, the smallest beam waist leads to the largest average energy gains, as expected.

Furthermore, in the case of the smallest waist size, 
i.e., $\Delta w_0 = 13 \lambda_0$, increasing the power of the laser by two orders of magnitude 
leads about $5$ times increase in net average energy gain, i.e., 
$\Delta E^{\gamma}_{10PW} \approx 5\times \Delta E^{\gamma}_{0.1PW}$.
On the other hand if we compare the average energy gained from the 
$P_0 = 10$ PW laser in case we increase the beam waist about $\sqrt{10}$ times to 
$\Delta w_0=41\lambda_0$, the results change only about a factor of $2$,
as a function of the waist size, i.e., $\Delta E^{\gamma}_{13\lambda_0} \approx 2.3\times 
\Delta E^{\gamma}_{41\lambda_0}$.
In this case the relatively large waist also leads to a decrease in the energy 
gain from the less powerful laser, i.e., $P_0=0.1$ PW, by more than an order in magnitude.

In all cases, the average energy gain is about a few MeV at most, 
while the maximum energy gain could be about $20-30$ times of this average for the highest intensity.
For peak intensities above $10^{19}$ $\textrm{W/cm}^2$ corresponding to a 0.1 PW laser,
electrons with kinetic energy of a few MeV were reported in Refs. \cite{He_2003,He_2004}, 
in agreement with our results.
For intensities above $10^{21}$ $\textrm{W/cm}^2$, relatively few electrons that are initially found 
in the close vicinity of the longitudinal axis, where the intensity is largest, 
the net energy gain may reach $\Delta E^{\gamma} \simeq 100-160$ MeV. 
This energy gain is comparable to a few times $m_e c^2 a_0$, while for $a_0 \geq 5$ the peak 
energy gain is $\Delta E^{\gamma} \geq m_e c^2 a_{0}$.
Note that for the $P_0=10$ PW laser, the dimensionless transverse coordinates leading to largest energy are
$x_{0,d} = 0.677$ and $y_{0,d}=0.484$ for $\Delta w_0 = 13\lambda_0$. 
These values are scaled by $23/13$ and $41/13$ for the larger spots.

We stress here again, that these results are practically insensitive to the polarization of the laser. 
For the largest part of the randomly distributed electrons there are no observable differences 
between the mean or the maximum energy gained due to the polarization of the laser. 
Of course, the maximum energy gained temporarily during the 
interaction might be larger in case of linear polarization due to the difference in intensity, field, and 
electron dynamics but that might not be reflected in the exit energies after scattering.

We also note that around the very center of the Gaussian pulse 
in the close vicinity of the intensity maximum the differences between CP and LP 
become visible especially for the smallest beam waist as apparent 
in Fig. \ref{fig:Energy_CPLP}c. 
On the other hand, we have verified that for an already very small radius of $r_0=0.15w_0$ 
that contains about $5\%$ of laser power 
transmitted at the highest intensity, the differences between the energy gains are only about a few MeV. 
The larger the radius around the intensity peak the smaller the average intensity while 
the differences between CP and LP are diminished.
This is clearly reflected even for the highest laser power where for the larger beam waists 
no difference between CP and LP pulses are observed.

\begin{figure}[hbt!]
\vspace{-0.2cm} 
\includegraphics[width=6.cm, height=4.5cm]
{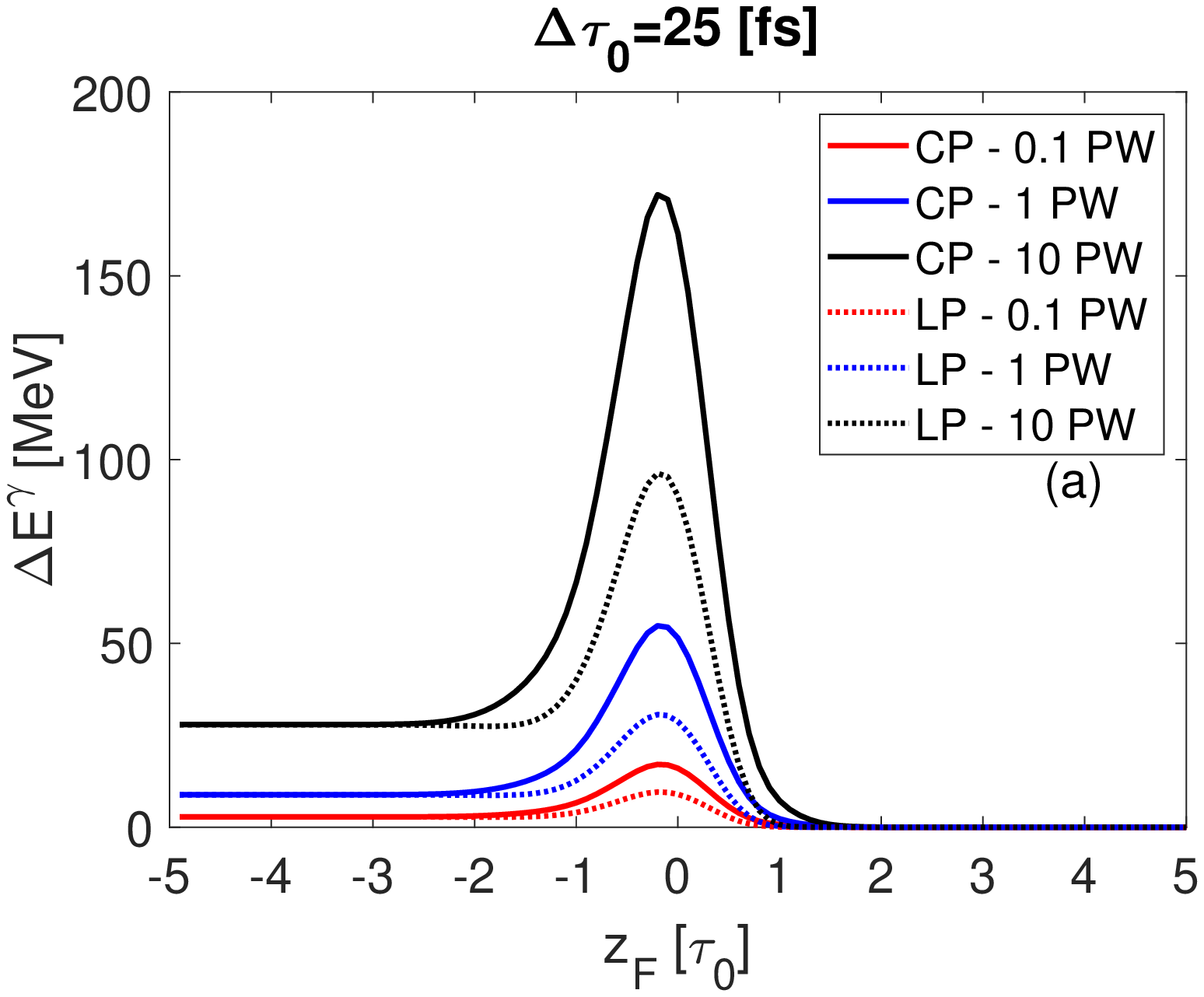}\\
\includegraphics[width=6.cm, height=4.5cm]
{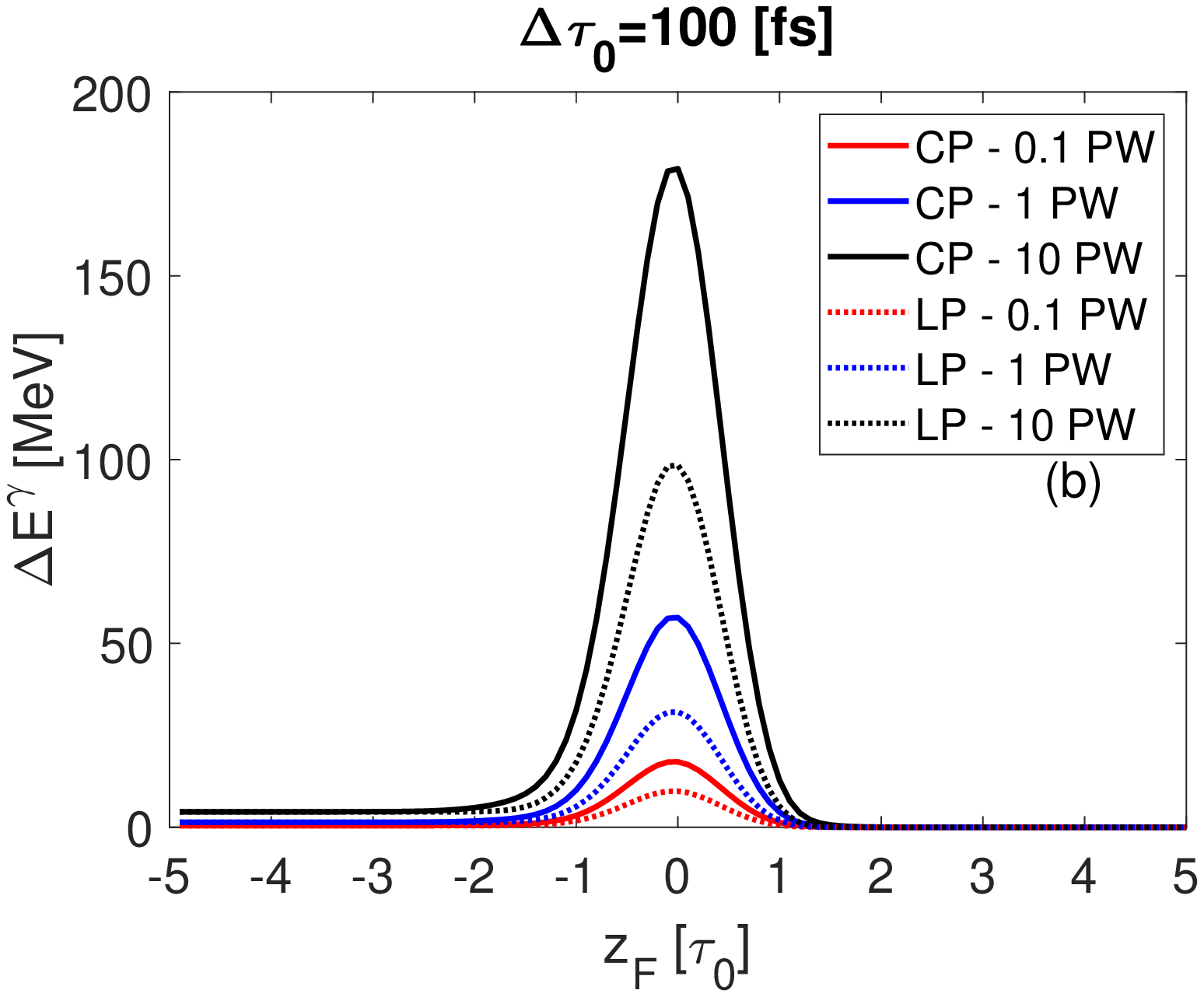}
\caption{The energy gain of an electron initially found at $z_0 =0$ 
and $x_0=y_0\equiv r_0/2$ as a function of the pulse peak position $z_F$. 
(a) The figure on the top corresponds to a $\Delta \tau_0 = 25$ fs pulse duration.
(b) The figure on the bottom for $\Delta \tau_0 = 100$ fs pulse duration.
The red, blue and black dotted and full lines correspond to initial waist radii 
of $\Delta w_{0}= \left\{13, 23, 41\right\}\lambda _{0}$ 
and laser powers of $P_0=\left\{0.1, 1, 10\right\}$ PW respectively. 
The dotted lines represent energy gain in linearly while the full lines in 
circularly polarized Gaussian beams.}
\label{fig:zF0_scan}
\end{figure}

\begin{figure*}[hbt!]
\vspace{-0.2cm} 
\includegraphics[width=16.4cm, height=4.5cm]
{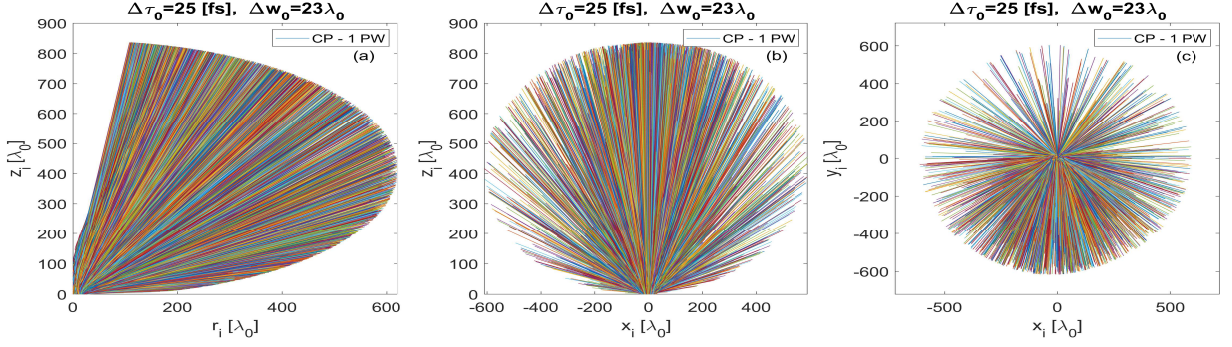}
\caption{The same as Fig. \ref{fig:XYZ_Gauss00} but for $z_F=0$.
(a) The electron trajectories in coordinate space. 
(b) The longitudinal trajectories $z_i$, as function of the radial distance $r_i$, 
and as a function of $x_i$. 
(c) The trajectories in the transverse plane $\left(x_i,y_i\right)$.}
\label{fig:XYZ_Gauss00_zF0}
\end{figure*}

\begin{figure*}[hbt!]
\vspace{-0.2cm} 
\includegraphics[width=16.4cm, height=4.5cm]
{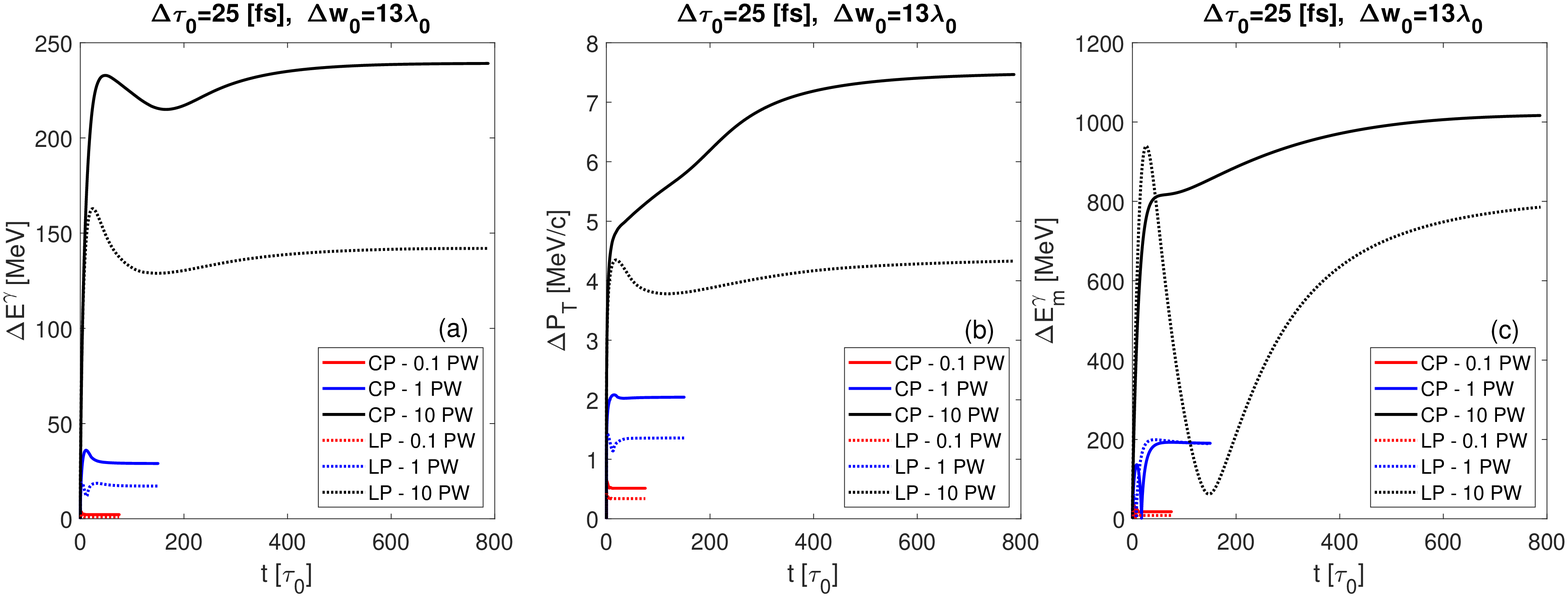} 
\includegraphics[width=16.4cm, height=4.5cm]
{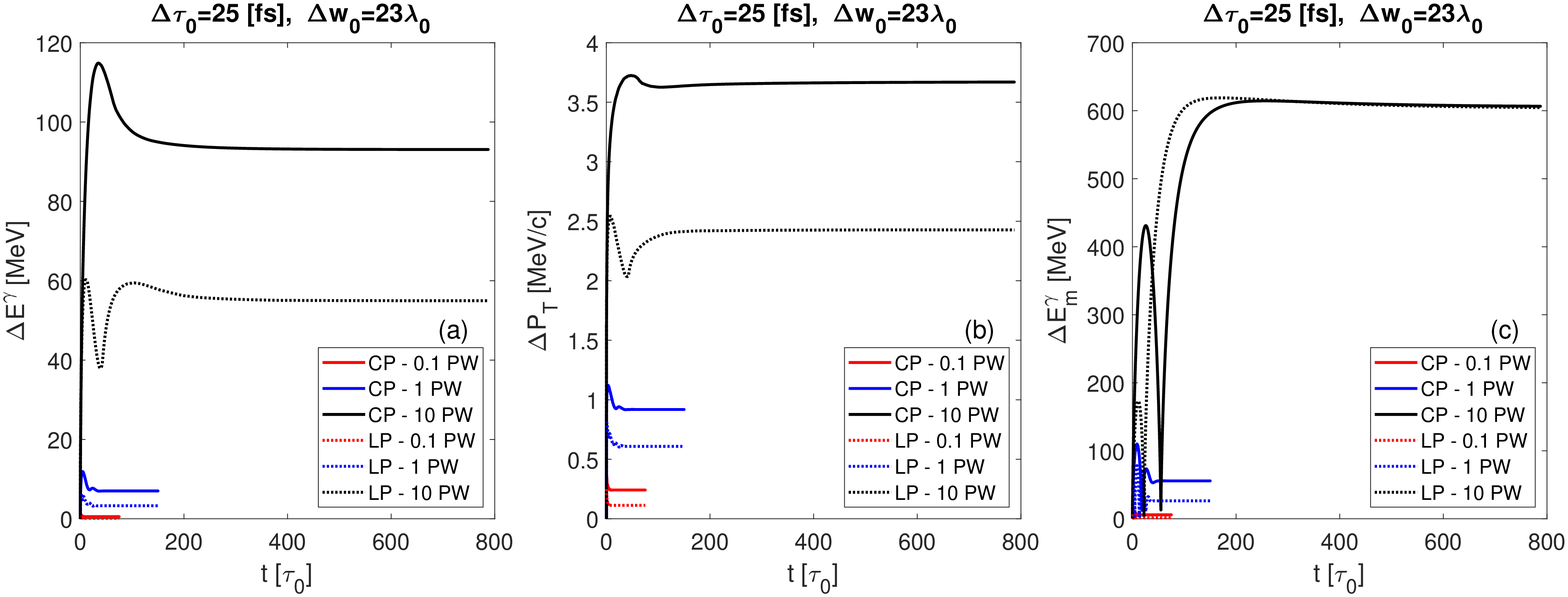} 
\includegraphics[width=16.4cm, height=4.5cm]
{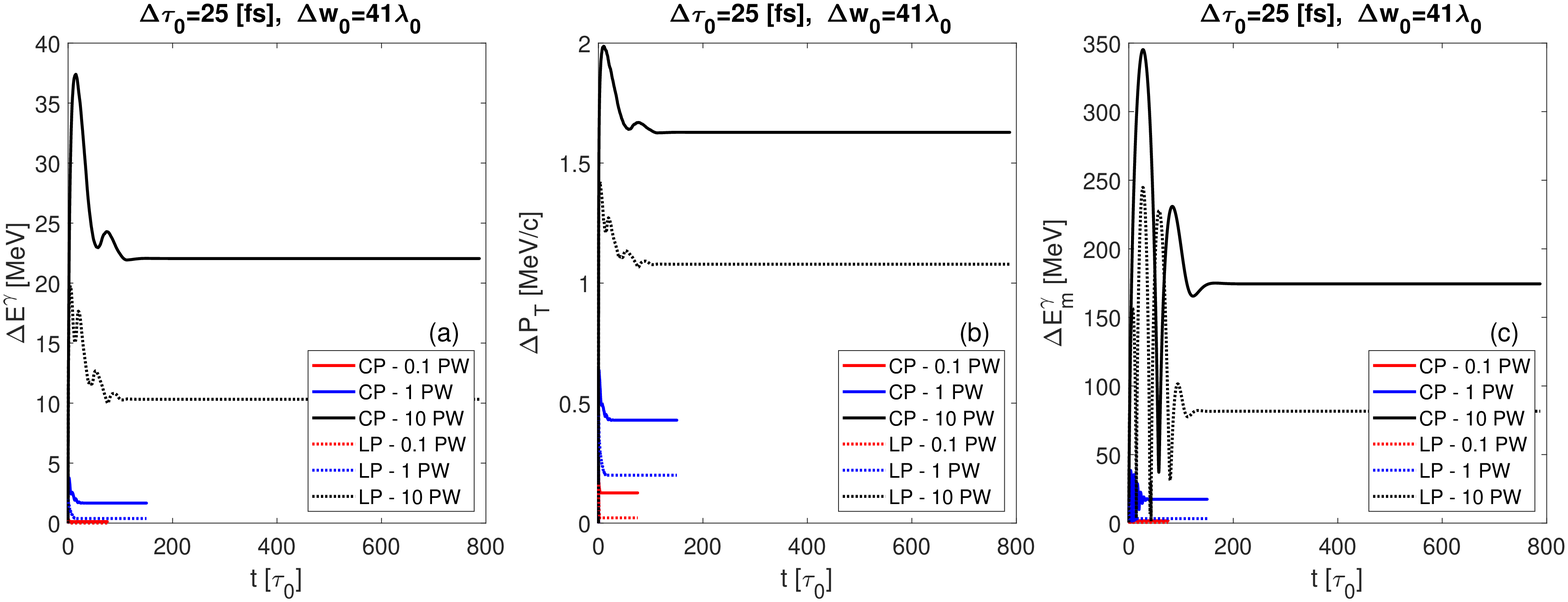}
\caption{The same as Fig. \ref{fig:Energy_CPLP}, except for $z_F=0$.
Upper row: with initial beam waist of $\Delta w_0=13\lambda_0$. 
(a) The time evolution of the mean net energy gain of electrons for laser powers 
of $P_0 = \left\{0.1, 1, 10\right\}$PW, 
with red, blue and black correspondingly. The full and dotted lines 
are for CP and LP pulses respectively.
(b) the average transverse momentum gained from the pulse as a function of time.
(c) the time evolution of the highest energy electron corresponding 
to given laser power. 
Middle row: with initial beam waist of $\Delta w_0=23\lambda_0$.
Lower row: with initial beam waist of $\Delta w_0=41\lambda_0$.
}
\label{fig:Energy_CPLP_zF0}
\end{figure*}

\begin{figure}[hbt!]
\vspace{-0.2cm} 
\includegraphics[width=8.2cm, height=4.5cm]
{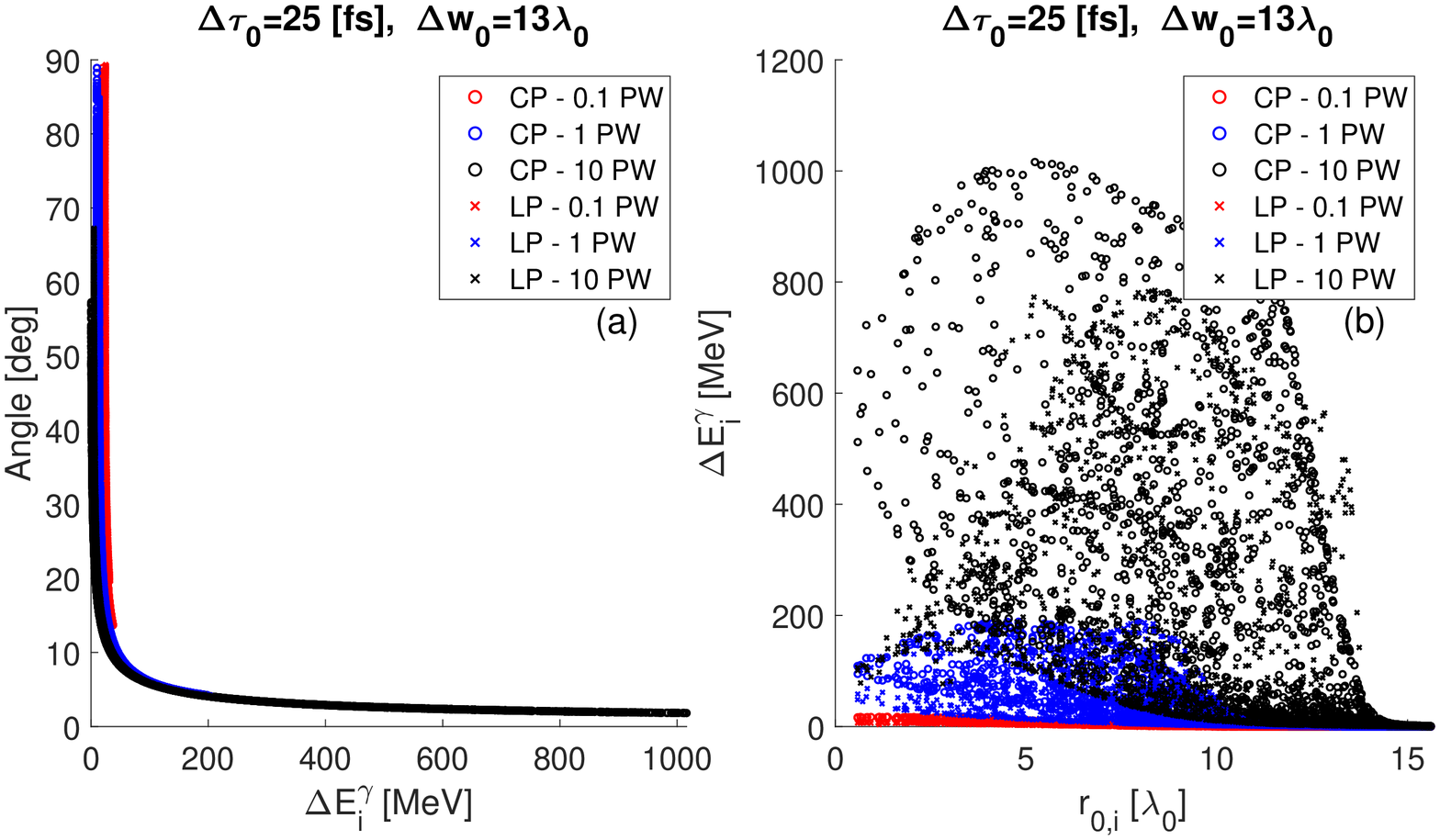} \\
\includegraphics[width=8.2cm, height=4.5cm]
{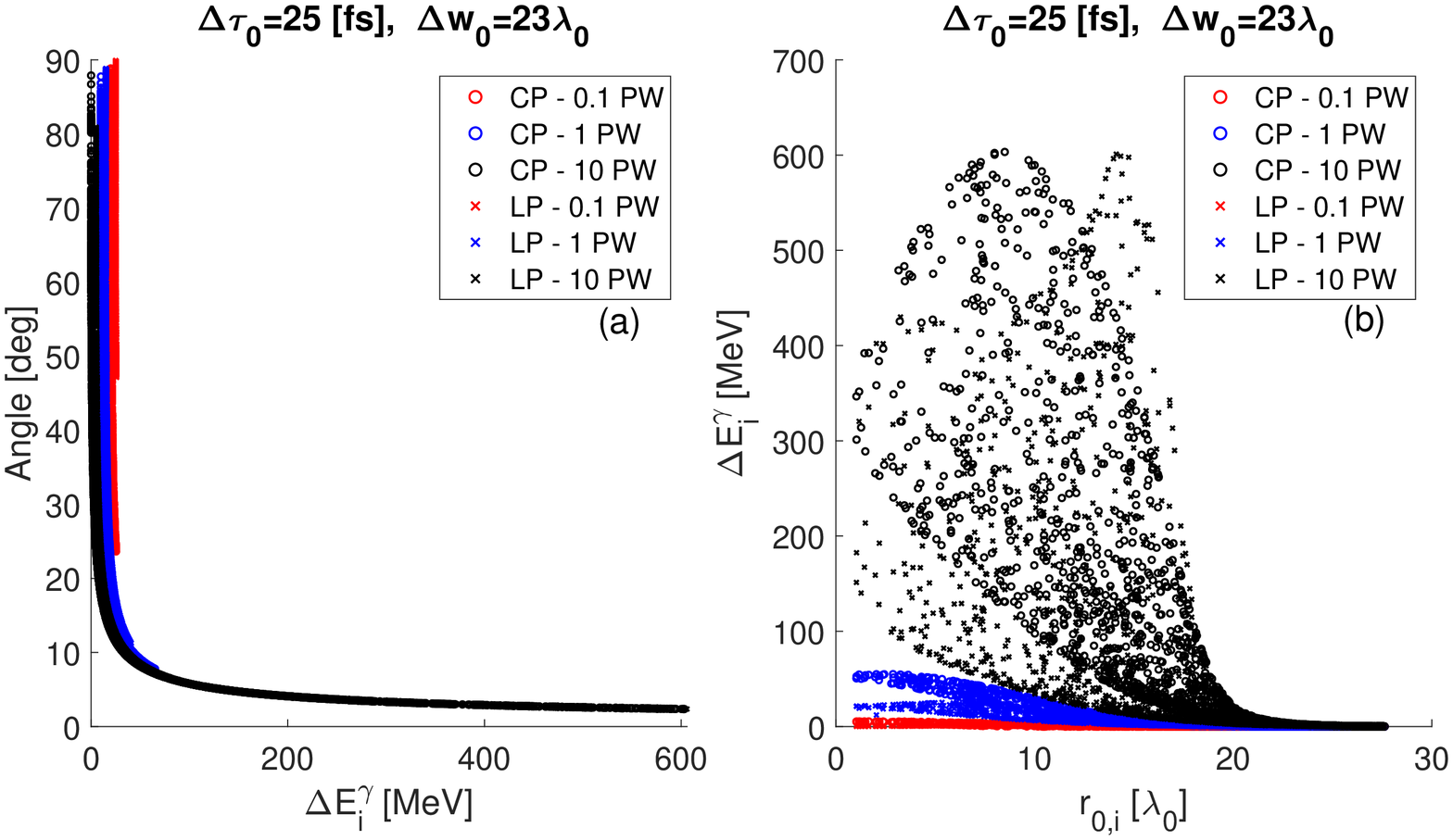} \\
\includegraphics[width=8.2cm, height=4.5cm]
{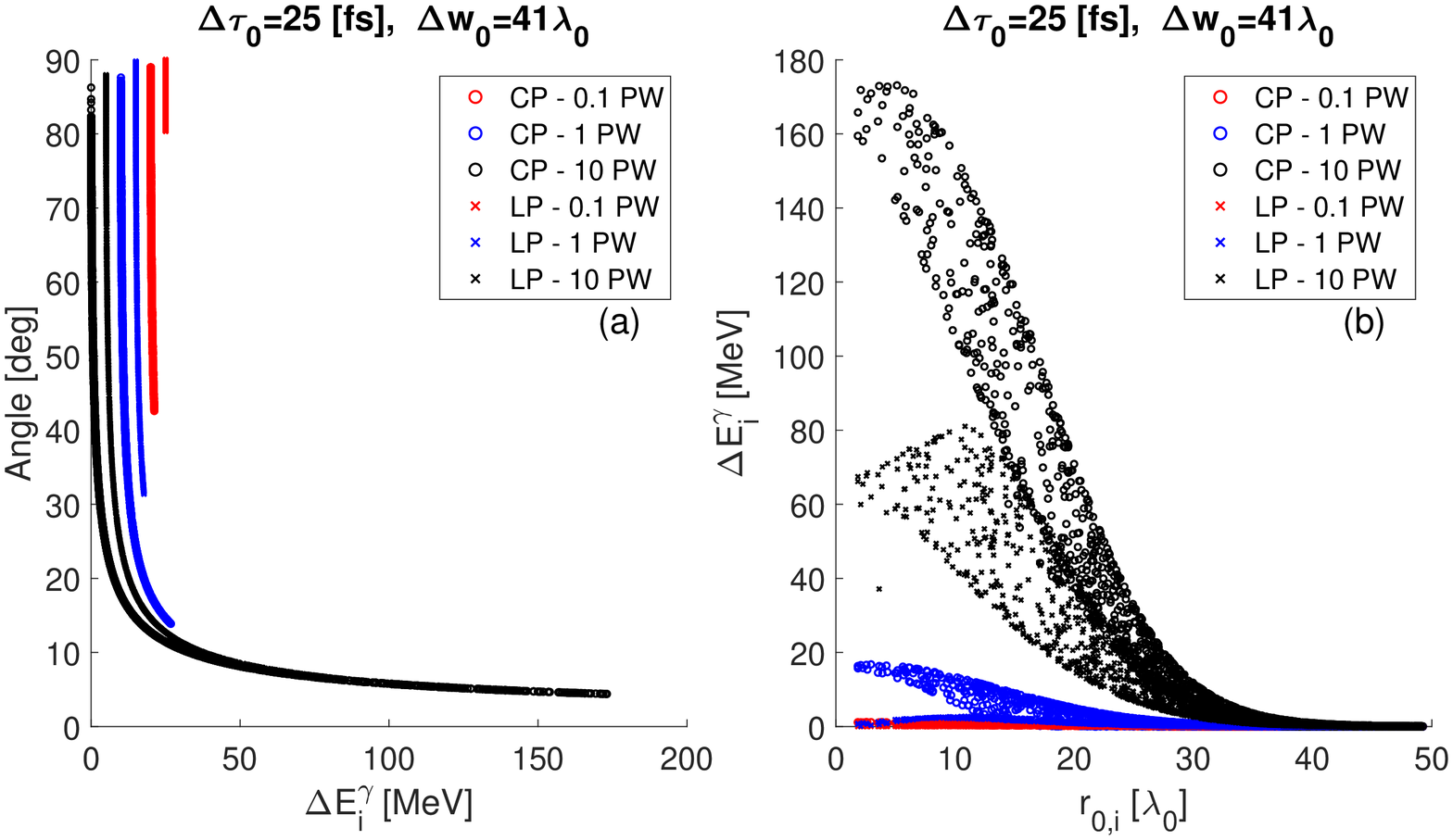}
\caption{The same as Fig. \ref{fig:Angular_CPLP}, except for $z_F=0$.
(a) The scattering angle as a function of energy gain. 
The data points are shifted for better visibility, otherwise they lay on the same curve.
(b) The energy gained form the pulse as a function of 
the initial radial distribution of electrons.
Middle row: with initial beam waist of $\Delta w_0 = 23 \lambda$.
Lower row: with initial beam waist of $\Delta w_0 = 41 \lambda$.
}
\label{fig:Angular_CPLP_zF0}
\end{figure}

\subsection{Partial and half pulse interaction $z_F = 0$}

In this section we discuss the effect of shifting the position of the initial 
intensity peak $z_F$ or the focus of the laser pulse while keeping the electrons 
at $z_{0}=0$ on the optical axis. 
The shift of the intensity peak corresponds to changing the interaction domain of 
the laser pulse \cite{Pandari_2018}, 
and therefore we can investigate the dependence of energy gain 
on the temporal envelope and the longitudinal position of the focus of the pulse. 

It is well known that in a symmetric pulse the energy gained from the front or 
the first half of pulse is gradually lost in the other half of the pulse throughout 
the optical cycles. 
Therefore by gradually removing the first part of the pulse, the electrons will not lose 
but retain a larger net energy. 
This is essentially what happens when a prepulse or the front of the pulse reaches and 
ionizes the matter while leaving behind free electrons at rest. 
Therefore such electrons are not captured by the whole pulse and hence their 
evolution is different.

For this study, we have kept the optimal parameters as listed in Table \ref{Power_table}, 
but now we have fixed the initial position of electrons at $x_0=y_0 \equiv w_0/2$.
Using this setup, the longitudinal position of the peak of the pulse was varied between 
$z_F = \left(-5\tau_0c, 5\tau_0c\right)$, and we have plotted the net energy 
gained as a function of $z_F$. 

The outcome of this analysis is shown in Fig. \ref{fig:zF0_scan} for different 
laser powers and polarizations.  
We observe that starting about $z_{F} > - 2\tau_0$ and approaching $z_{F}\simeq 0$ 
we obtain larger and larger energy gains. This is due the fact that we are gradually 
removing the front part of the laser pulse, i.e., we modify the Gaussian time-envelope, 
see Eq. (\ref{time_envelope}). 
The temporal envelope, is centered at $z_{F} = 0$ and decreases with increasing 
$|z_F|$. Therefore larger initial field amplitudes lead to larger acceleration.

For $z_{F} \lesssim - 2\tau_0 c$, we observe that the energy gain is independent on 
the polarization, in accordance with our previous results.
However, in contrast to the previously studied case already for $z_{F} > - 2\tau_0 c$ 
the optical polarization of the pulse becomes important.
We found that for $z_{F} \approx 0$ a circularly polarized pulse may lead to 
almost two times larger energy gain than a linearly polarized pulse albeit 
the latter has a larger amplitude.

Similarly to the previous sections \ref{full_pulse} and \ref{full_pulse_optimized}, here
we have analyzed the case of half-pulse interaction using a pulse envelope with $z_{F} =0$. 
With such initial conditions, the particle trajectories and energy gains
have changed considerably.

The main difference between full-pulse and half-pulse interaction for circularly polarized 
Gaussian laser is a strongly anisotropic distribution in both the transverse and 
longitudinal directions.
There are more electrons scattered in the negative than in the 
positive $y$-direction, see the particle trajectories shown in 
Figs. \ref{fig:XYZ_Gauss00_zF0}b and \ref{fig:XYZ_Gauss00_zF0}c.
This is also true in case of a LP laser, however 
it is much less pronounced than in the former case.

Now, once again using the different cases listed in Table \ref{Power_table}, 
we show and compare the corresponding results in Figs. \ref{fig:Energy_CPLP_zF0} 
for circularly and linearly polarized Gaussian beams respectively.
The results obtained in case of half pulse interaction lead to the following observations.

The maximum net energy gained (for initially zero constant phases) 
obtained in half pulse interaction, can be almost an order of magnitude larger, i.e., 
$\Delta E^{\gamma} \simeq 1$ GeV, than from full pulse interaction as shown on 
Figs. \ref{fig:Energy_CPLP_zF0}.  
This maximum kinetic energy gain is in agreement with the theoretical approximation 
\cite{Stupakov_2000,Dodin_2003}, 
the so-called ponderomotive limit $\Delta E^\gamma \approx m_e c^2 a^2_0/2$,
and favors circularly polarized rather than linearly polarized pulses, hence it 
is best estimated by $a_{0,CP}$ instead of $a_{0,LP}$.
We also note here that within or beyond the ponderomotive limit, 
energy gain of GeV magnitude can also be realized in other various
regimes, see Refs. \cite{Arefiev_2013,Leblanc_2016,Salamin_2020}.
Note that, in contrast to full-pulse interactions, in half-pulse interactions 
the initial phase of electrons, i.e., position, is of importance for the energy gain, 
as shown in Figs. \ref{fig:Angular_CPLP_zF0}b. 

In conclusion the net energy gain increases with intensity and decreasing beam waist and 
the optimal value is given by the smallest waist for a given laser power.
Furthermore since the energy gained is also a function of the interaction area covered 
by the laser pulse, more net energy is retained when the electron interacts 
with smaller frontal area of the pulse.

\section{Conclusions}
\label{Conclusions}

In this paper we have studied direct laser acceleration of electrons by  
fundamental Gaussian pulses relevant for multi-PW femtosecond lasers at ELI-NP.

In the case of full pulse interactions we have estimated the optimal values of beam waist 
and we have found that the net energy gain increases with increasing laser power 
but at the same time it is largest for an optimal beam waist $\Delta w_0$. 
The optimal beam waists at full width at half maximum correspond to 
$\Delta w_0=\left\{13,23,41\right\}\lambda_0$ 
for laser powers of $P_0 = \left\{0.1,1,10\right\}$ PW. 
In our study we have obtained an average energy gain of a few, i.e., $2-6$ MeV and 
highest-energy electrons about $160$ MeV for the $10$ PW laser, for these optimal values.
On the other hand in the case of half-pulse interaction the average energy gain can 
be of the order of $100$'s MeV for the tightest waist of $\Delta w_0=13 \lambda_0$ 
with highest energy electrons of $1$ GeV for the $10$ PW laser.

An investigation of DLA with higher-order Laguerre-Gaussian 
modes will follow the present work.

\begin{acknowledgments}
The authors thank D. Doria, Z. Harman, K. Spohr, J. F. Ong and K. Tanaka for 
suggestions and corrections.
E.~Moln\'ar also thanks H. S. Ghotra for the comparison of early results 
and valuable discussions.
D. Stutman acknowledges support by a grant of 
Ministry of Education and Research, CNCS-UEFISCDI, project number 
PN-IIIP4-ID-PCCF-2016–0164, within PNCDI III.
The authors are thankful for financial support from the Nucleu Project PN 19060105. 
\end{acknowledgments}

\section*{Author contribution statement}
E.M. performed the calculations and the preparation of the manuscript. 
All the authors have read, supported and approved the final manuscript.  



\begin{thebibliography}{99}

\bibitem{Shimoda_1962}
K. Shimoda, 
Appl. Opt. 1, 33 (1962).

\bibitem{Tajima_1979}
T. Tajima and J. M. Dawson
Phys. Rev. Lett. 43, 267 (1979); https://doi.org/10.1103/PhysRevLett.43.267 

\bibitem{Scully_1991}
Marlan O. Scully and M. S. Zubairy
Phys. Rev. A 44, 2656 (1991); https://doi.org/10.1103/PhysRevA.44.2656

\bibitem{Sarachik_1970}
E.~S.~Sarachik and G.~T.~Schappert,
Phys.\ Rev.\ D \textbf{10}, 2738 (1970).

\bibitem{Malka:1997}
G. Malka, E. Lefebvre, and J. L. Miquel
Phys.\ Rev.\ Lett. \textbf{78} (1997), 3314;
https://doi.org/10.1103/PhysRevLett.78.3314


\bibitem{Moore_1995}
C.~I.~Moore, J.~P.~Knauer, and D.~D.~Meyerhofer,
Phys.\ Rev.\ Lett \textbf{74}, 2439 (1995).

\bibitem{Leemans_2006}
W. P. Leemans, B. Nagler, A. J. Gonsalves, Cs. Tóth, K. Nakamura, C. G. R. Geddes, E. Esarey, 
C. B. Schroeder and S. M. Hooker 
Nat. Phys. 2, 696 (2006).

\bibitem{Leemans_2014}
W. P. Leemans et al.,
Phys. Rev. Lett. 113, 245002; 
https://doi.org/10.1103/PhysRevLett.113.245002


\bibitem{Pukhov_1999}
A. Pukhov, Z.-M. Sheng, and J. Meyer-ter-Vehn,
Phys. Plasmas 6, 2847 (1999); https://doi.org/10.1063/1.873242

\bibitem{Mourou_2006}
Gerard A. Mourou, Toshiki Tajima, and Sergei V. Bulanov
Rev.\ Mod.\ Phys.\ textbf{78}, 309 (2006);
https://doi.org/10.1103/RevModPhys.78.309

\bibitem{Esarey_2009}
E. Esarey, C. B. Schroeder, and W. P. Leemans
Rev. Mod. Phys. 81, 1229 (2009);
https://doi.org/10.1103/RevModPhys.81.1229

\bibitem{Arefiev_2012}
Alexey V. Arefiev, Boris N. Breizman, Marius Schollmeier, and Vladimir N. Khudik
Phys. Rev. Lett. 108, 145004 (2012); https://doi.org/10.1103/PhysRevLett.108.145004


\bibitem{Arefiev_2016}
A.~V.~Arefiev, V.~N.~Khudik, A.~P.~L.~Robinson, G.~Shvets, L.~Willingale, and M.~Schollmeier
Phys. Plasmas 23, 056704 (2016); https://doi.org/10.1063/1.4946024

\bibitem{Arefiev_2019}
Tianhong Wang, Vladimir Khudik, Alexey Arefiev, and Gennady Shvets 
Phys. Plasmas 26, 083101 (2019);  https://doi.org/10.1063/1.5110407

\bibitem{Tanaka_2020}
K. A. Tanaka, K. M. Spohr, D. L. Balabanski, S. Balascuta, L. Capponi, M. O. Cernaianu, 
M. Cuciuc, A. Cucoanes, I. Dancus, A. Dhal, B. Diaconescu, D. Doria, P. Ghenuche, D. G. Ghita, 
S. Kisyov, V. Nastasa, J. F. Ong, F. Rotaru, D. Sangwan, P.-A. Söderström, D. Stutman, 
G. Suliman, O. Tesileanu, L. Tudor, N. Tsoneva, C. A. Ur, D. Ursescu, and N. V. Zamfir 
Matter and Radiation at Extremes 5, 024402 (2020); 
https://doi.org/10.1063/1.5093535 

\bibitem{He_2003}
Feng He, Wei Yu, Peixiang Lu, Han Xu, Liejia Qian, Baifei Shen, Xiao Yuan, Ruxin Li and Zhizhan Xu,
Phys.\ Rev.\ E \textbf{68}, 046407 (2003).

\bibitem{He_2004}
He Feng, Yu Wei, Lu Peixiang and Xu Han
Plasma Sci. Technolgy 6 2492 (2004); http://iopscience.iop.org/1009-0630/6/5/013


\bibitem{Kalashnikov_2015}
M. Kalashnikov, A. Andreev, K. Ivanov, A. Galkin, V. Korobkin, M. Romanovsky, O. Shiryaev, M. Schnuerer, J. Braenzel, V. Trofimov 
Laser and Particle Beams, 33(3), 361-366 (2015);
doi:10.1017/S0263034615000403

\bibitem{Esarey_1995}
E.~Esarey, P.~Sprangle and J.~Krall,
Phys.\ Rev.\ E \textbf{52}, 5443 (1995).


\bibitem{Hartemann_1995}
F.~V.~Hartemann, S.~N.~Fochs, G.~P.~Le Sage, N.~C.~Luhmann, Jr., J.~G.~Woodworth, 
M.~D.~Perry, Y.~J.~Chen and A.~K.~Kerman,
Phys.\ Rev.\ E \textbf{51}, 4833 (1995).

\bibitem{Hartemann_1998}
F.~V.~Hartemann, J.~R.~Van Meter, A.~L.~Troha, E.~C.~Landahl, N.~C.~Luhmann, Jr., J.~G.~Woodworth, 
H.~A.~Baldis, Atul.~Gupta and A.~K.~Kerman,
Phys.\ Rev.\ E \textbf{58}, 5001 (1998).


\bibitem{Quesnel_1998}
B.~Quesnel and P.~Mora,
Phys.\ Rev.\ E \textbf{58}, 3791 (1998).


\bibitem{Stupakov_2000}
G.~V.~Stupakov and M.~.S.~Zolotorev, 
Phys.\ Rev.\ Lett \textbf{86}, 5274 (2000).


\bibitem{Wang_2000}
P.~X.~Wang, Y.~K.~Ho, X.~Q.~Yuan, Q.~Kong, N.~Cao, L.~Shao, A.~M.~Sessler, 
E.~Esarey, E.~Moshkovich, Y.~Nishida, N.~Yugami, H.~Ito, J.~X.~Wang, and S.~Scheid
Journal of Applied Physics \textbf{91}, 856 (2002); https://doi.org/10.1063/1.1423394


\bibitem{Maltsev_2002}
A. Maltsev and T. Ditmire
Phys.\ Rev.\ Lett. \textbf{90}, 053002 (2002);
https://doi.org/10.1103/PhysRevLett.90.053002


\bibitem{Salamin_2002}
Yousef I. Salamin and Christoph H. Keitel
Phys.\ Rev.\ Lett. \textbf{88}, 095005; 
https://doi.org/10.1103/PhysRevLett.88.095005


\bibitem{Dodin_2003}
I.~Y.~Dodin and N.~J.~Fisch
Phys.\ Rev.\ E \textbf{68}, 056402 (2003); https://doi.org/10.1103/PhysRevE.68.056402

\bibitem{He_2005}
He Feng, Yu Wei, Lu Peixiang, Xu Han, Shen Baifei, Li Ruxin and Xu Zhizhan,
Plasma Science and Technology, Vol.7, No. 4, 2968 (2005).


\bibitem{Salamin_2006}
Yousef I.~Salamin
Phys.\ Rev.\ A \textbf{73}, 043402 (2006); 
https://doi.org/10.1103/PhysRevA.73.043402

\bibitem{Gupta_2007}
Devki Nandan Gupta, Niti Kant, Dong Eon Kim, and Hyyong Suk
Phys. Lett. A 368 (2007) 402–407,
https://doi.org/10.1016/j.physleta.2007.04.030

\bibitem{Galkin_2008}
A. L. Galkin, V. V. Korobkin, M. Yu. Romanovsky, and O. B. Shiryaev
Phys. Plasmas 15, 023104 (2008);
https://doi.org/10.1063/1.2839349

\bibitem{Harman_2008}
Yousef I. Salamin, Zolt\'an Harman, and Christoph H. Keitel
Phys.\ Rev.\ Lett. \textbf{100}, 155004 (2008);
https://doi.org/10.1103/PhysRevLett.100.155004

\bibitem{Harman_2011}
Zolt\'an Harman, Yousef I. Salamin, Benjamin J. Galow, and Christoph H. Keitel
Phys. Rev. A 84, 053814 (2011); https://doi.org/10.1103/PhysRevA.84.053814

\bibitem{Fortin_2010}
Pierre-Louis Fortin, Michel Pieche and Charles Varin,
J.\ Phys.\ B:\ At.\ Mol.\ Opt \textbf{43}, 025401 (2010).


\bibitem{Arefiev_2013}
A. P. L. Robinson, A. V. Arefiev, and D. Neely
Phys.\ Rev.\ Lett.\  \textbf{111}, 065002 (2013);
https://doi.org/10.1103/PhysRevLett.111.065002

\bibitem{Harjit_2017}
Harjit Singh Ghotra and Niti Kant
Optics Communications 383 (2017) 169–176;
https://doi.org/10.1016/j.optcom.2016.08.061

\bibitem{Kant_2020}
Niti Kant, Jyoti Rajput and Arvinder Singh
Eur. Phys. J. D 74: 142 (2020);
https://doi.org/10.1140/epjd/e2020-100241-y


\bibitem{Jackson_Book_1999} J.~D.~Jackson, 
\textit{Classical Electrodynamics}, Third Edition, 
John Wiley \& Sons, New York, (1999).


\bibitem{Siegman_Lasers_1986} A.~E.~Siegman, 
\textit{Lasers}, 
University Science Books; Revised ed. edition (1986).

\bibitem{Goldsmith_book_1998} P.~F.~Goldsmith,
\textit{Quasioptical Systems: Gaussian Beam Quasioptical Propogation and Applications}
Wiley-IEEE Press (1998).


\bibitem{Kawata_2011}
Y. Y. Li, Y. J. Gu, Z. Zhu, X. F. Li, H. Y. Ban, Q. Kong, and S. Kawata
Phys. Plasmas 18, 053104 (2011); https://doi.org/10.1063/1.3581062


\bibitem{Cicchitelli_1990}
Lorenzo Cicchitelli, H. Hora, and R. Postle
Phys. Rev. A 41, 3727 (1990);
https://doi.org/10.1103/PhysRevA.41.3727

\bibitem{Pandari_2018}
M. Rezaei-Pandari, M. Akhyani, F. Jahangiri, A.R. Niknam, R. Massudi
Optics Communications 429 (2018) 46–52;
https://doi.org/10.1016/j.optcom.2018.07.081

\bibitem{Leblanc_2016}
Th\'evenet, M., Leblanc, A., Kahaly, S. et al. 
Nature Phys 12, 355–360 (2016); 
https://doi.org/10.1038/nphys3597
 
\bibitem{Salamin_2020}
Meng Wen, Yousef I. Salamin, and Christoph H. Keitel
Phys. Rev. Applied 13, 034001; https://doi.org/10.1103/PhysRevApplied.13.034001


\end{thebibliography}
\end{document}